\documentclass[a4paper,12pt]{article}

\usepackage{mathrsfs}%hanamoji
\usepackage{graphicx}
\usepackage{bm}
\usepackage{amssymb}
\usepackage{amsmath}
\usepackage{array}
 \usepackage{cite}

\begin{document}

%%%%%
%\begin{minipage}{16cm}

\title{Information Thermodynamics: \\ Maxwell's Demon in Nonequilibrium Dynamics}
\author{Takahiro Sagawa and Masahito Ueda}

\maketitle

\begin{abstract}
We review theory of information thermodynamics which incorporates effects of measurement and feedback into nonequilibrium thermodynamics of a small system, and discuss how the second law of thermodynamics should be extended for such situations. We address the issue of the maximum work that can be extracted from the system in the presence of a feedback controller (Maxwell's demon) and provide a few illustrative examples. We also review a recent experiment that realized a Maxwell's demon based on a feedback-controlled ratchet.
\end{abstract}

\tableofcontents

\section{Introduction}

The profound interrelationship between information and thermodynamics was first brought to light by J.~C.~Maxwell~\cite{Maxwell} in his gedankenexperiment about a hypothetical being of intelligence which was later christened by William Thomson as Maxwell's demon.  Since then, numerous discussions have been spurred as to whether and how Maxwell's demon is compatible with the second law of thermodynamics~\cite{Demon,Maruyama,Maroney0}. 

To understand the roles of Maxwell's demon, let us consider a situation in which a gas is confined in a box surrounded by adiabatic walls.  A barrier is inserted at the center of the box to divide it into two.  The temperatures of the gases in the two boxes are assumed to be initially the same.  We assume that the demon is present near the barrier and can close or open a small hole in the barrier.
If a faster-than-average (slower-than-average) molecule comes from the right (left) box, the demon opens the hole.   Otherwise, the demon keeps it closed.  By doing so over and over again, the temperature of the left gas becomes higher than that of the right one, in apparent contradiction with the second law of thermodynamics. 
This example illustrates the two essential roles of the demon:
\begin{itemize}
\item The demon observes individual molecules, and obtains the information about their velocities.
\item The demon opens or closes the hole based on each measurement outcome, which is the feedback control.  
\end{itemize}
In general,  feedback control implies that a  control protocol depends on the measurement outcome, or equivalently, we control a system based on the obtained information~\cite{Doyle,Astrom}.
The crucial point here is that the measurements are performed at the level of thermal fluctuations (i.e., the demon can distinguish the velocities of individual molecules).  Therefore, Maxwell's demon can be characterized as a feedback controller that utilizes information about a thermodynamic system at the level of thermal fluctuations (see also Fig.~1).  
\begin{figure}[htbp]
 \begin{center}
 \includegraphics[width=70mm]{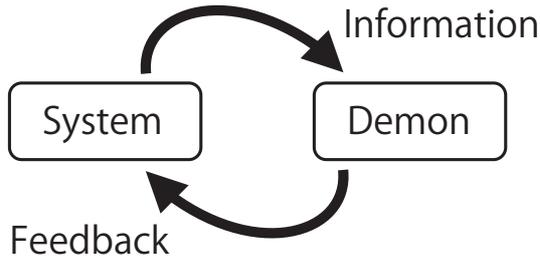}
 \end{center}
 \caption{Maxwell's demon as a feedback controller. The demon performs feedback control based on the information obtained from measurement at the level of thermal fluctuations.} 
\end{figure}

In the nineteenth century, it was impossible to observe and control individual atoms and molecules, and therefore it was not necessary to take into account the effect of feedback control in the formulation of  thermodynamics.  However, due to the recent advances in manipulating  microscopic systems, the effect of feedback control on thermodynamic systems has become relevant to real experiments.
We can simulate the role of Maxwell's demon in real experiments and can reduce the entropy of small thermodynamic systems.  

In this chapter, we review a general theory of thermodynamics that involves measurements and feedback control~\cite{Lloyd1,Caves,Lloyd2,Nielsen,Touchette,Zurek1,Kieu,Allahverdyan,Touchette2,Quan,Cao1,Kim0,Kim,Lopez,Allahverdyan2,Sagawa-Ueda1,Jacobs,Cao2,Touchette3,Sagawa-Ueda3,Ponmurugan,Suzuki,Horowitz1,Toyabe3,SWKim,Morikuni,Ito,Horowitz2,Abreu,Jarzynski5,Sagawa,Dong,Pekola,Horowitz3,Granger,Lahiri,Lu,Sagawa-Ueda4}.  
We generalize the second law of thermodynamics by including information contents concerning the thermodynamics of feedback control.
%because the conventional thermodynamics was not intended to describe feedback-controlled processes. 
%We identify the information contents that play crucial roles in the thermodynamics of feedback control.  
We note that, by the ``demon,'' we mean a type of devices that perform feedback control at the level of thermal fluctuations.

This chapter is organized as follows.  In Sec.~2, we discuss the Szilard engine which is a prototypical model of Maxwell's demon and examine the consistency between the demon and the second law.   In Sec.~3, we review information contents that are used in the following sections. In Sec.~4, we discuss a generalized second law of thermodynamics with feedback control, which is the main part of this chapter.  In Sec.~5, we generalize nonequilibrium equalities such as the fluctuation theorem and the Jarzynski equality to the case with feedback control.  In Sec.~6, we discuss the energy cost (work) that is needed for  measurement and  information erasure.  In Sec.~7, we conclude this chapter.

%%%%%%%%%%%%%%%%%%%%%%%%%%%%%%%%%%%%%%%%%%%%%%%%%%%

\section{Szilard Engine}

In 1929, L.~Szilard proposed a simple model of  Maxwell's demon that illustrates the quantitative relationship between information and thermodynamics~\cite{Szilard}.  In this section, we briefly review the model, which is called the Szilard engine, and discuss its physical implications.

The Szilard engine consists of  a single-particle gas that is in contact with a single heat bath at temperature $T$. 
By a measurement, we obtain one bit of information about the position of the particle and use that information to extract work from the engine via feedback control.
While the engine eventually returns to the initial equilibrium, the total amount of the extracted  work is positive.  The details of the control protocol are as follows (see Fig.~2).

\begin{figure}[htbp]
 \begin{center}
 \includegraphics[width=70mm]{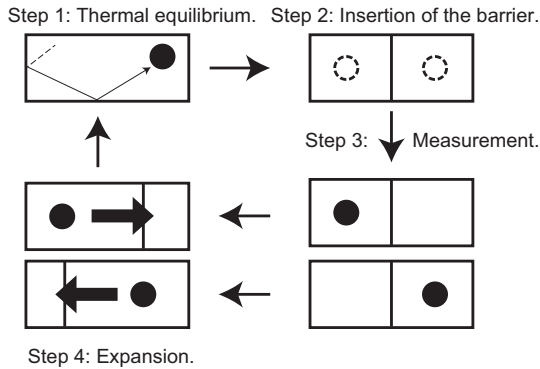}
 \end{center}
 \caption{The Szilard engine.  See the text for details.}   
\end{figure}

\

\textit{Step 1: Initial state.}  We prepare a single-particle gas in a box of volume $V_0$, which is at  thermal equilibrium with temperature $T$.

\

\textit{Step 2: Insertion of the barrier.}  We insert a barrier in the middle of the box, and divide it into two with equal volume $V_0/2$.

\

\textit{Step 3: Measurement.} We perform an error-free measurement of the position of the particle to find which box the molecule is in.  
Because the particle will be found to be in each box with probability $1/2$, the amount of information gained from this measurement is one bit.  We note that one ($= \log_2 2$) bit of information in the binary logarithm corresponds to $\ln 2$ nat of information in the natural logarithm.

\

\textit{Step 4: Feedback.}  If the particle is in the left (right) box, we quasi-statically expand it by moving the barrier to the rightmost (leftmost) position.  
By this process, we can extract  work $W_{\rm ext}$  given by
\begin{equation}
W_{\rm ext} = \int_{V_0/2}^{V_0} \frac{k_{\rm B}T}{V}dV = k_{\rm B}T \ln 2,
\end{equation}
where we used  $pV = k_{\rm B}T$ with $p$,  $V$, and $k_{\rm B}$ respectively being the pressure, the volume and the Boltzmann constant.  This process corresponds to feedback control, because the direction of the expansion (i.e., left or right) depends on the measurement outcome.  After this expansion,  the gas returns to the initial thermal equilibrium with volume $V_0$.

\

%The total amount of work that we extract during this process is positive.  
The extracted work is proportional to the obtained information with proportionality constant $k_{\rm B}T$.  This is due to the fact that the entropy of the system is effectively decreased by $ \ln 2$ via feedback control, and the decrease in entropy leads to the increase in the free energy by $k_{\rm B}T \ln 2$, which is the resource of the extracted work.  

The Szilard engine \textit{prima facie} seems to contradict the second law of thermodynamics, which dictates that one cannot extract positive work from a single heat bath with a thermodynamic cycle (Kelvin's principle).  
In fact, the Szilard engine is consistent with the second law, due to an additional energy cost (work) that is needed for the measurement and information erasure of the measurement device or the demon.   This additional cost compensates for the excess work extracted from the engine.  In his original paper, Szilard argued that there must be an entropic cost for the measurement process~\cite{Szilard}. We stress that the work performed on the demon need not be transferred to the engine; only the information obtained by the measurement should be utilized for the feedback. This is the crucial characteristic of the information heat engine.

By utilizing the obtained information in feedback control, we can extract work from the engine without decreasing its free energy, or we can increase the engine's free energy without injecting any work to the engine directly.  The resource of the work or the free energy is thermal fluctuations of the heat bath; by utilizing information via feedback, we can rectify the thermal energy of the bath and convert it into the work or the free energy.   This method allows us to control the energy balance of the engine beyond the conventional thermodynamics.
We shall call such a feedback-controlled heat engine as an ``information heat engine.''  
The Szilard engine works as the simplest model that illustrates the quintessence of information heat engines. 
Quantum versions of the Szilard engine have also been studied~\cite{Zurek1,SWKim,Dong,Lu}.
%In the following, we develop a general theory that connects information and thermodynamics  by using the concepts of information theory.   

%%%%%%%%%%%%%%%%%%%%%%%%%%%%%%%%%%%%%%%%%

\section{Information Content in Thermodynamics}

In this section, we briefly review the Shannon information and the mutual information~\cite{Cover-Thomas,Shannon}.  In particular, the mutual information plays a crucial role in thermodynamics of information processing.  
%In Sec.~3.3, we will discuss two typical examples.

\subsection{Shannon Information}

Let $x \in X$ be a probability variable which represents a finite set of possible events. We write as $P[x]$ the probability  of  event $x$ being realized.
The information content that is associated with event $x$ is then defined as $- \ln P[x]$, which implies that the rarer an event is, the more information is associated with it.  
The Shannon information is then given by the average of $- \ln P[x]$ over all possible events:
\begin{equation}
H(X) := - \sum_x P[x] \ln P[x].
\label{Shannon}
\end{equation}
The Shannon information satisfies $0 \leq H(X) \leq \ln N$,
where $N$ is the number of the possible events (the size of set $X$).
Here, the lower bound ($H(X) = 0$) is achieved  if $P[x] = 1$ holds for some $x$; in this case, the event is indeed deterministic, while the upper bound ($H(X) = \ln N$) is achieved  if $P[x] = 1/N$ for arbitrary  $x$.
In general, the Shannon information characterizes the randomness of a probability variable; the more random the variable, the greater the Shannon information. 
Consider, for example, a simple case in which $x$ takes two values: $x=0$ or $x=1$.  We set $P[0] =: p$ and $P[1] =: 1-p$ with $0 \leq p \leq 1$.  The Shannon information is then given by $H(X) = - p \ln p - (1-p) \ln (1-p)$,
which  takes the maximum value $\ln 2$ for $p=1/2$ and the minimal value $0$ for $p = 0$ or  $1$.

\subsection{Mutual Information}

The mutual information characterizes the correlation between two probability variables.   Let $x \in X$ and $y \in Y$ be the two probability variables, and $P[x,y]$ be their joint distribution.  
The marginal distributions are given by $P[x] = \sum_y P[x,y]$, $P[y] = \sum_x P[x,y]$.
If the two variables are statistically independent, then $P[x,y] = P[x]P[y]$.  Otherwise,  they are correlated.
If the two variables are perfectly correlated, the joint distribution satisfies 
\begin{equation}
P[x,y] = \delta (x,f(y))P[x] =  \delta (x, f(y))P[y],
\label{perfect_correlation}
\end{equation}
where $\delta (\cdot, \cdot)$ is the Kronecker's delta and $f(\cdot)$ is a bijection function on $Y$.  
For example, if  $P[0,1] = P[1,0] = 1/2$ and $P[0,0] = P[1,1] = 0$ with $X = \{ 0,1 \}$ and $Y= \{ 0, 1 \}$, the two variables are perfectly correlated with $f(0) = 1$ and $f(1) = 0$.
If $f(\cdot)$ is the identity function that satisfies $f(y) = y$ for any $y$, Eq.~(\ref{perfect_correlation}) reduces to $P[x,y] = \delta (x,y)P[x] =  \delta (x, y)P[y]$.

The conditional probability of $x$ for given $y$ is given by $P[x|y] =P[x,y] / P[y]$.
If  the two probability  variables are statistically independent, the conditional probability reduces to $P[x|y] = P[x]$.  
This implies that we cannot obtain any information about $x$ from knowledge of $y$.
On the other hand, in the case of the perfect correlation (\ref{perfect_correlation}), we obtain $P[x|y] = \delta (x, f(y))$.  This means that we can precisely estimate $x$ from $y$ by $x=f(y)$.

%We note that the Bayes' theorem can straightforwardly be obtained from the foregoing arguments.  By noting that $P[x,y] = P[y|x] P[x]$ and $P[y] = \sum_x P[y|x] P[x]$ hold, we obtain the Bayes' theorem 
%\begin{equation}
%P[x|y] = \frac{P[y|x] P[x]}{\sum_x P[y|x] P[x]},
%\label{Bayes}
%\end{equation}
%which enables us to calculate conditional probability $P[x|y]$ only from $P[y|x]$ and $P[x]$.

We next introduce the joint Shannon information and the conditional Shannon information.  The Shannon information for the joint probability $P[x,y]$ is given by
\begin{equation}
H(X,Y) := - \sum_{xy} P[x,y] \ln P[x,y].
\end{equation}
On the other hand, the Shannon information for the conditional probability $P[x | y]$, where $x$ is the relevant probability variable,  is given by
\begin{equation}
H(X|y) := - \sum_x P[x|y] \ln P[x|y].
\end{equation}
By averaging $H(X|y)$ over $y$, we define the conditional Shannon information
\begin{equation}
H(X|Y) :=  \sum_y P[y] H(X|y) = - \sum_{xy} P[x,y] \ln P[x|y].
\label{conditional1}
\end{equation}
The conditional Shannon information satisfies the following properties:
\begin{equation}
H(X|Y) = H(X,Y) - H(Y), \ H(Y|X) = H(X,Y) - H(X).
\label{conditional2}
\end{equation}
By definition, $H(X|y) \geq 0$ and $H(X|Y) \geq 0$. Hence
\begin{equation}
H(X,Y) \geq H(Y), \ H(X,Y) \geq H(X),
\label{inequality0}
\end{equation}
which implies that the randomness decreases if only one of the two variables is concerned.  

The mutual information is defined by
\begin{equation}
I(X:Y) := H(X) + H(Y) - H(X,Y),
\label{mutual1}
\end{equation}
or equivalently,
\begin{equation}
I(X:Y) = \sum_{xy} P[x,y] \ln \frac{P[x,y]}{P[x]P[y]}.
\label{mutual2}
\end{equation}
As shown below, the mutual information satisfies
\begin{equation}
0 \leq I(X:Y) \leq H(X), \ 0 \leq I(X:Y) \leq H(Y). 
\label{inequality_mutual}
\end{equation}
Here, $I(X:Y) = 0$ is achieved if $X$ and $Y$ are statistically independent, i.e., $P[x,y] = P[x]P[y]$.  On the other hand, $I(X:Y) = H(X)$ is achieved if $H(X|Y) = 0$, or equivalently, if $H(X|y) = 0$ for any $y$.  This condition is equivalent to the condition that, for any $y$, there exists a single $x$ such that $P[x | y] = 1$, which implies that we can estimate $x$ from $y$ with certainty.    Similarly, $I(X:Y) = H(Y)$ is achieved if $H(Y|X) = 0$.  In particular, if the correlation between $x$ and $y$ is perfect such that Eq.~(\ref{perfect_correlation}) holds, $I(X:Y) = H(X) = H(Y)$.  In general, the mutual information describes the correlation between two probability variables;  the more strongly $x$ and $y$ are correlated, the larger $I(X:Y)$ is.

The proof of inequalities~(\ref{inequality_mutual}) goes as follows.
Because $\ln (t^{-1}) \geq 1- t$ for $t>0$, where the equality is achieved if and only if $t=1$,  we obtain
\begin{equation}
- \sum_x P[x,y] \ln \frac{P[x]P[y]}{P[x,y]} \geq \sum_x P[x,y] \left( 1 - \frac{P[x]P[y]}{P[x,y]} \right) = 0,
\end{equation} 
which implies $I(X:Y) \geq 0$.  On the other hand, Eq.~(\ref{mutual1}) leads to
\begin{equation}
I(X:Y) = H(X) - H(X|Y) = H(Y) - H(Y|X),
\label{mutual3}
\end{equation}
which implies $I(X:Y) \leq H(X)$ and  $I(X:Y) \leq H(Y)$.

We note that Eqs.~(\ref{conditional2}), (\ref{mutual1}), and (\ref{mutual3}) can be illustrated by using a Venn diagram shown in Fig.~3~\cite{Cover-Thomas}.  This diagram is useful to memorize the relationship among $H(X|Y)$, $H(Y|X)$, and $I(X:Y)$. 
\begin{figure}[htbp]
 \begin{center}
 \includegraphics[width=50mm]{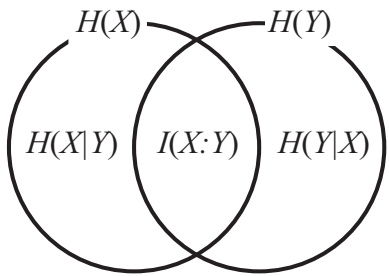}
 \end{center}
 \caption{A Venn diagram~\cite{Cover-Thomas} illustrating the relationship among different information contents.  The entire region  represents the joint Shannon information $H(X,Y).$} 
\end{figure}

The mutual information can be used to characterize the effective information that can be obtained by measurements.
Let us consider a situation in which $x$ is a phase-space point of a physical system and $y$ is an outcome that is obtained from a measurement on the system.  In the case of the Szilard engine, $x$ specifies the location of the particle (``left'' or ``right''), and $y$ is the measurement outcome.  If the measurement is error-free as we assumed in Sec.~2, $x=y$ is always satisfied and the correlation between the two variables is perfect.  In this case, $I(X:Y) = H(X) = H(Y)$ holds, and therefore the obtained information can be characterized by the Shannon information as is the argument in Sec.~2.  
In general, there exist measurement errors, and the obtained information by the measurement needs to be characterized by the mutual information.  The less the amount of the measurement error is, the more the mutual information is.

We next discuss the cases in which the probability variables take continuous values.  The probability distributions such as $P[x,y]$ and $P[x]$ should then be interpreted as probability densities, where the corresponding probabilities are given by $P[x,y]dxdy$ and $P[x]dx$ with $dxdy$ and $dx$  being the integral elements.  The Shannon information of $x$ can be formally defined as 
\begin{equation}
H(X) := - \int dx P[x] \ln P[x].
\label{Shannon_continuous}
\end{equation}
However, Eq.~(\ref{Shannon_continuous}) is not invariant under the transformation of the variable.  In fact, if we change $x$ to $x'$ such that $P[x]dx = P[x']dx'$, Eq.~(\ref{Shannon_continuous}) is given by
\begin{equation}
H(X) = - \int dx' P[x'] \ln P[x'] - \int dx' P[x'] \ln \Bigl| \frac{dx'}{dx} \Bigr|.
\end{equation}
Thus, that  the Shannon information is not uniquely defined for the case of continuous variables.  Only when we fix some probability variable, we can  give the Shannon information a unique meaning. 
On the other hand, the mutual information is defined as
\begin{equation}
I(X:Y) := \int dxdy P[x,y] \ln \frac{P[x,y]}{P[x]P[y]},
\end{equation}
which is invariant under the transformation of the variables.  In this sense, the mutual information is uniquely defined for the cases of continuous variables, regardless of the choices of probability variables.

\subsection{Examples}

We now discuss two typical examples of probability variables:  discrete and continuous variables.

\paragraph{Example 1 (Binary channel)}
We consider a binary channel with which at most one bit of information is sent from variable $x$ to $y$ [see Fig.~4 (a)].  
Let $x= 0, 1$ be the sender's bit and $y= 0, 1$ be the receiver's bit.
We regard this binary channel as a model of a measurement, in which $x$ describes the state of the measured system and $y$ describes the measurement outcome.
We assume that the error in the communication (or the measurement error) is characterized by 
\begin{equation}
\begin{split}
P[x=0 | y=0]  = 1-\varepsilon_0, \ P[x=0 | y=1]  = \varepsilon_0, \\
P[x=1 | y=0]  = \varepsilon_1, \ P[x=1 | y=1]  = 1-\varepsilon_1,
\end{split}
\end{equation}
where $\varepsilon_0$ and $\varepsilon_1$ are the error rates for $x=0$ and $x=1$, respectively.  The crucial assumption here is that the error property is only characterized  by a pair $(\varepsilon_0, \varepsilon_1)$, which is independent of the probability distribution of $x$. 
If $\varepsilon_0 = \varepsilon_1 =: \varepsilon$, this model is called a binary symmetric channel. 
\begin{figure}[htbp]
 \begin{center}
 \includegraphics[width=100mm]{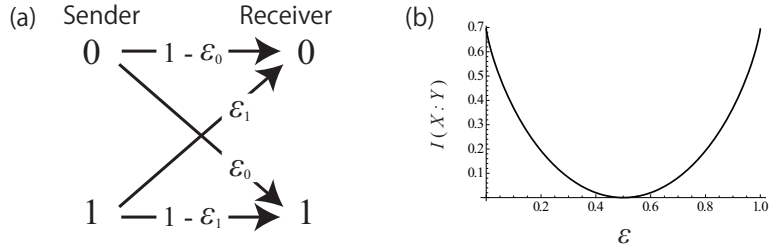}
 \end{center}
 \caption{(a) A binary channel with error rates $\varepsilon_0$ and $\varepsilon_1$.  (b) Mutual information $I(X:Y)$ versus error rate $\varepsilon$ for a binary symmetric channel with $\varepsilon_0 = \varepsilon_1 =: \varepsilon$ and $p=1/2$, which gives $I(X:Y) = \ln 2 + \varepsilon \ln \varepsilon + (1-\varepsilon) \ln (1-\varepsilon)$.} 
\end{figure}

Let $P[x=0] =: p$ and  $P[x=1] =: 1-p$ be the probability distribution of  $x$.
The joint distribution of $x$ and $y$ is then given by
$P[x=0, y=0]  = p(1-\varepsilon_0)$, $P[x=0, y=1]  = p\varepsilon_0$, $P[x=1, y=0]  = (1-p)\varepsilon_1$, $P[x=1, y=1]  = (1-p)(1-\varepsilon_1)$, and the distribution of  $y$ is given by
$P[y=0] = p(1-\varepsilon_0) + (1-p)\varepsilon_1 =: q$, $P[y=1] = p \varepsilon_0 + (1-p)(1-\varepsilon_1) =: 1-q$.  By definition, we can show that the mutual information is given by
\begin{equation}
I(X:Y) = H(Y) - p H(\varepsilon_0) - (1-p) H (\varepsilon_1),
\label{mutual_binary}
\end{equation}
where $H(Y) := -q \ln q - (1-q) \ln (1-q)$ is the Shannon information for $Y$,  and we defined $H(\varepsilon_i) := - \varepsilon_i \ln \varepsilon_i - (1 - \varepsilon_i) \ln (1- \varepsilon_i)$ for $i=0$ and $1$.
From Eq.~(\ref{mutual_binary}), $I(X:Y) = H(Y)$ holds for $(\varepsilon_0, \varepsilon_1) = (0,0)$, $(0,1)$, $(1,0)$, or $(1,1)$. For  a binary symmetric channel,  Eq.~(\ref{mutual_binary}) reduces to $I(X:Y) = H(Y) -  H(\varepsilon)$.
Figure 4 (b) shows  $I(X:Y)$ versus $\varepsilon$ for the case of $p=1/2$.
The mutual information takes the maximum value if  $\varepsilon = 0$ or $\varepsilon = 1$.  In this case, we can precisely estimate $x$ from $y$.  We note that, if $\varepsilon = 1$, we can just relabel ``$0$'' and ``$1$'' of $y$ such that $x=y$ holds.

\paragraph{Example 2 (Gaussian channel)}

We next consider a Gaussian channel with continuous variables $x$ and $y$.  Let $x$ be the sender's variable or the ``signal'' and $y$ be the receiver's variable or the ``outcome.''  We can also interpret that $x$ describes the phase-space point of a physical system such as the position of a Brownian particle, and that $y$ describes the outcome of the measurement on the system.
We assume that the error is characterized by a Gaussian noise
\begin{equation}
P[y | x] = \frac{1}{\sqrt{2 \pi N}} \exp \left( - \frac{(y-x)^2}{2N} \right),
\label{Gauss_error}
\end{equation}
where $N$ is the variance of the noise.  
For simplicity, we also assume that the probability density of $x$ is also Gaussian:
\begin{equation}
P[x] = \frac{1}{\sqrt{2\pi S}} \exp \left( - \frac{x^2}{2S} \right),
\label{Gauss_x}
\end{equation}
where $S$ is the variance of $x$.  The joint probability density of $x$ and $y$ is then given by 
\begin{equation}
P[x,y] = P[y|x]P[x] = \frac{1}{\sqrt{4\pi^2 SN}} \exp \left( - \frac{x^2}{2S} - \frac{(y-x)^2}{2N} \right).
\label{Gauss_xy}
\end{equation}
On the other hand, the probability density of $y$ is given by
\begin{equation}
P[y] =  \frac{1}{\sqrt{2 \pi (S+N)}} \exp \left( - \frac{y^2}{2(S+N)} \right),
\label{Gauss_y}
\end{equation}
which implies that the variance of the outcome $y$ is enhanced by factor $1+N/S$ compared with the original variance $S$ of the signal $x$.
We can also calculate  the conditional probability density as 
\begin{equation}
P[x|y] = \frac{P[x,y]}{P[y]} = \frac{1}{\sqrt{2 \pi SN / (S+N)}} \exp \left( - \frac{S+N}{2SN} \left( x - \frac{S}{S+N}y \right)^2 \right),
\label{Gauss_conditional}
\end{equation}
which implies that the variance of the conditional distribution of $x$ is suppressed compared with the original variance $S$ by a factor of $1+S/N$.

We can straightforwardly calculate the mutual information as
\begin{equation}
I(X:Y) = \frac{1}{2} \ln \left( 1 + \frac{S}{N} \right),
\label{Gauss_mutual}
\end{equation}
which is determined by the signal-to-noise ratio $S/N$ alone.  In the limit of $S/N \to 0$, where  the noise dominates the signal, the mutual information vanishes.
On the other hand, in the limit of $S/N \to \infty$ where the noise is negligible compared with the signal, the mutual information diverges, as the variable is continuous.

%%%%%%%%%%%%%%%%%%%%%%%%%%%%%%%%%%%%%%%%%%%%%%%%%%%%%%

\section{Second Law of Thermodynamics with Feedback Control}

In this section, we discuss a universal upper bound of the work that can be extracted from information heat engines such as the Szilard engine.   Starting from a general argument for isothermal processes in Sec.~4.1, we discuss two models with which the universal bound is achieved in Sec.~4.2 and 4.3.  Moreover, we will discuss an experimental result that demonstrates an information heat engine in Sec.~4.4.  We will also discuss the Carnot efficiency with two heat baths in Sec.~4.5.

\subsection{General Bound}

In the conventional thermodynamics, we extract a  work from a heat engine by changing external parameters such as the volume of a gas or the frequency of an optical tweezer.  In addition to such parameter changes, we perform measurements and feedback control for the case of information heat engines.

Suppose that we have a thermodynamic engine that is in contact with a single heat bath at temperature $T$.  We perform a measurement on a thermodynamic engine, and obtain $I$ of mutual information.  After that, we  extract a positive amount of work by changing external parameters.  The crucial point here is that the protocol of changing the external parameters can depend on the measurement outcome via feedback control.

In the case of the Szilard engine, we obtain $\ln 2$ nat of information, and extract $k_{\rm B}T \ln 2$ of work.  %We can then ask if we can extract a larger amount of work than $k_{\rm B}T \ln 2$ when we select a better feedback protocol than the Szilard engine.
How much work can we  extract in principle under the condition that we have $I$ of mutual information about the system? 
The answer of this fundamental question is given by the following inequality~\cite{Sagawa-Ueda1,Sagawa-Ueda3}:
\begin{equation}
W_{\rm ext} \leq -\Delta F + k_{\rm B}TI,
\label{second_feedback}
\end{equation}
where $W_{\rm ext}$ is the average of the work that is extracted from the engine,  and $\Delta F$ is the free-energy difference of the engine between the initial and final states.
Inequality~(\ref{second_feedback}) has been proved for both quantum and classical regimes~\cite{Sagawa-Ueda1,Sagawa-Ueda3}.  
However, the mutual information in (\ref{second_feedback}) needs to be replaced by a quantum extension of the mutual information~\cite{Sagawa-Ueda1,Groenewold,Ozawa} for quantum cases.
We will prove inequality~(\ref{second_feedback}) for classical cases by invoking the detailed fluctuation theorem in Sec.~5.

Inequality~(\ref{second_feedback}) states that we can extract an excess work up to $k_{\rm B}TI$ if we utilize $I$ of mutual information obtained by the measurement.
%Inequality~(\ref{second_feedback}) sets the fundamental upper bound of the work that can be extracted from a single heat bath with the assistance of feedback control. 
In the conventional thermodynamics, the upper bound of the extractable work is bounded only by the free-energy difference $\Delta F$, which is determined by the initial and final values of external parameters.  For information heat engines, the mutual information  is also needed to determine the upper bound of the extractable work.
In this sense, inequality~(\ref{second_feedback}) is a generalization of the second law of thermodynamics for feedback-controlled processes,  in which thermodynamic variables ($W$ and $\Delta F$) and the information content ($I$) are treated on an equal footing.

The equality in (\ref{second_feedback}) is achieved with the ``best'' protocol, which means that the process is quasi-static and the post-feedback state is independent of  the measurement outcome, i.e., we utilize all the obtained information.
This condition is achieved by the Szilard engine, as discussed above.
Some models that achieves the equality in (\ref{second_feedback}) have been proposed~\cite{Jacobs,Horowitz2,Abreu,Sagawa-Ueda4}.  Two of them~\cite{Abreu,Sagawa-Ueda4} are discussed in Secs.~4.2 and 4.3.
The Szilard engine, which gives $W = k_{\rm B}T \ln 2$,  $\Delta F = 0$, and  $I = \ln 2$, achieves the upper bound of the extractable work, and its special role in information heat engines parallels that of the Carnot cycle in  conventional thermodynamics.

\subsection{Generalized Szilard Engine}

We discuss a generalization of the Szilard engine with measurement errors and imperfect feedback~\cite{Sagawa-Ueda4}, with which the equality  in (\ref{second_feedback}) is achieved.  The control protocol is described  as follows (see Fig.~5):
\begin{figure}[htbp]
 \begin{center}
 \includegraphics[width=70mm]{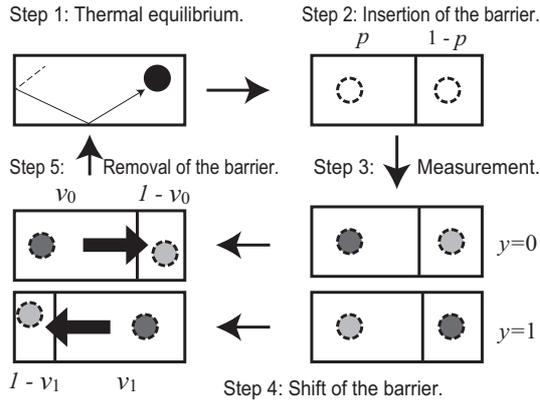}
 \end{center}
 \caption{Generalized Szilard engine.  See the text for details.} 
\end{figure}

\

\textit{Step 1: Initial state.}
A single-particle gas is in thermal equilibrium, which is in contact with a single heat bath at temperature $T$.

\

\textit{Step 2: Insertion of the barrier.}
We insert a barrier to the box and divide it into two with the volume ratio being $p : 1-p$.

\

\textit{Step 3: Measurement.}
We perform a measurement to find out which box the particle is in.  The possible outcomes are ``left'' and ``right,'' which we respectively denote as ``$0$'' and ``$1$.''
The measurement can then be modeled by the binary channel with error rates $\varepsilon_0$ and $\varepsilon_1$ (see Sec.~3), where $x$ ($= 0, 1$) specifies the location of the particle and $y$ ($=0, 1$) shows the measurement outcome.

\

\textit{Step 4: Feedback.}
We quasi-statically move the barrier depending on the measurement outcome.  If the outcome is ``left'' ($y=0$),  we move the barrier, so that the final ratio of the volumes of  the two boxes is given by  $v_0 : 1 - v_0$ ($0 \leq v_0 \leq 1$).  If the outcome is ``right'' ($y=1$),  we move the  the barrier, so that the final ratio of the volumes of  the two boxes is given by  $1- v_1 :  v_1$ ($0 \leq v_1 \leq 1$).  We note that the feedback protocol is  specified by $(v_0, v_1)$.

\

\textit{Step 5: Removal of the barrier.}
We remove the barrier, and the system returns to the initial thermal equilibrium by a free expansion.

\

%We note that the original Szilard engine corresponds to a special case with $\varepsilon_0 = \varepsilon_1 = 0$ and $v_0 = v_1 = 1$.
We now calculate the amount of the work that is extracted in \textit{step 4}.   By using the equation of states $pV = k_{\rm B}T$, we find that the extracted work is $k_{\rm B}T\ln [v_0 / p]$ for $(x,y) = (0,0)$, $k_{\rm B}T\ln [(1-v_1)/p]$ for $(0,1)$, $k_{\rm B}T \ln [(1-v_0)/ (1-p)]$ for $(1,0)$, and $k_{\rm B}T \ln [v_1 / (1-p)]$ for $(1,1)$.  Therefore the average work is given by 
\begin{equation}
\frac{W_{\rm ext}}{k_{\rm B}T} = p(1- \varepsilon_0 ) \ln \frac{v_0}{p} + p \varepsilon_0 \ln \frac{1-v_1}{p} + (1-p)\varepsilon_1 \ln \frac{1 - v_0}{1-p} + (1-p)(1-\varepsilon_1) \ln \frac{v_1}{1-p}.
\end{equation}

The mutual information obtained by the measurement is given by Eq.~(\ref{mutual_binary}).
The upper bound of inequality (\ref{second_feedback}) is not  necessarily achieved with a general  feedback protocol   $(v_0, v_1)$.
We then maximize $W_{\rm ext}$ in terms of $v_0$ and $v_1$.  The optimal feedback protocol with the maximum work is determined by equations $\partial W_{\rm ext} / \partial v_0 = 0$ and  $\partial W_{\rm ext} / \partial v_1 = 0$, which lead to $v_0 = p(1-\varepsilon_0) / q = P[x=0 | y=0]$ and  $v_1 = (1-p)(1-\varepsilon_1) / (1-q) = P[x=1 | y=1]$.
Therefore, we obtain the maximum work as
\begin{equation}
W_{\rm ext} = k_{\rm B}T I,
\label{Szilard_maximum}
\end{equation}
which achieves the upper bound of the generalized second law~(\ref{second_feedback}).

\subsection{Overdamped Langevin System}

We next discuss a feedback protocol for an overdamped Langevin system, which also achieves the upper bound of inequality~(\ref{second_feedback}) as shown in Ref.~\cite{Abreu}.    We consider a Brownian particle with a harmonic potential, which obeys the following overdamped Langevin equation:
\begin{equation}
\eta \frac{dx}{dt} =  -\lambda_1 (x - \lambda_2) + \xi (t),
\end{equation}
where $\eta$ is a friction constant and $\xi (t)$ is a Gaussian white noise satisfying $\langle \xi (t) \xi (t') \rangle = 2\eta k_{\rm B}T\delta (t-t')$ with $\delta (\cdot)$ being the delta function.
The harmonic potential can be controlled through two external parameters $(\lambda_1, \lambda_2)$ such that
\begin{equation}
V(x, \lambda_1, \lambda_2) = \frac{\lambda_1}{2} (x-\lambda_2 )^2, 
\end{equation}
where $\lambda_1$ and $\lambda_2$ respectively describe the spring constant and the center of the potential.
We consider the following feedback protocol (see Fig.~6).
\begin{figure}[htbp]
 \begin{center}
 \includegraphics[width=88mm]{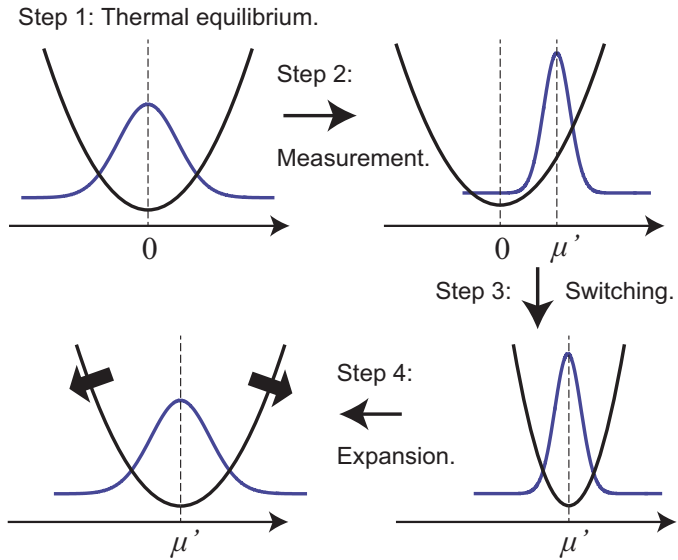}
 \end{center}
 \caption{Feedback control on a Langevin system with a harmonic potential.  See the text for details.} 
\end{figure}

\

\textit{Step 1: Initial state.} The particle is initially in thermal equilibrium with initial external parameters $\lambda_1 (0) =: k$ and $\lambda_2 (0) = 0$.

\

\textit{Step 2: Measurement.}  We measure the position of the particle and obtain outcome $y$.  The measurement error is assumed to be  Gaussian that is given by Eq.~(\ref{Gauss_error}), where $S$ is the variance of $x$ in the initial equilibrium state (i.e., $S = k_{\rm B}T / k$), and $N$ is the variance of the noise in the measurement [see Eq.~(\ref{Gauss_error})].  Immediately after the measurement, the conditional probability is given by Eq.~(\ref{Gauss_conditional}).  The obtained mutual information $I$ is given by Eq.~ (\ref{Gauss_mutual}).

\

\textit{Step 3: Feedback.}  Immediately after the measurement, we instantaneously change $\lambda_1$ from $k$ to $k' := (1+S/N)k$ and  $\lambda_2$ from $0$ to $\mu_y := Sy/(S+N)$.  By this change, the conditional distribution (\ref{Gauss_conditional}) becomes thermally equilibrated with new parameters $(\lambda_1, \lambda_2) = (k', \mu_y)$.

\

\textit{Step 4: Work extraction.} We quasi-statically expand the potential by changing $\lambda_1$ from $k'$ to $k$ thereby extracting the work.   The system then get thermally equilibrated with parameters $(\lambda_1, \lambda_2) = (k, \mu_y)$.

\

We now calculate the work that can be extracted from this engine.  Let $P[x,t | y]$ be the probability distribution of $x$ at time $t$ under the condition of $y$.
The average of the work for \textit{step 3} is given by
\begin{equation}
W_{\rm ext}^{(3)} =  \int\left[ V(x, k, 0) - V(x, k', \mu_y) \right]  P[x,0 | y ] P[y]  dxdy  = 0,
\label{work3}
\end{equation}
where $P[x,0 | y ]$ is given by $P[x|y]$ in Eq.~(\ref{Gauss_conditional}), and $P[y]$  is given by Eq.~(\ref{Gauss_y}).   
The work for \textit{Step 4} is given by
\begin{equation}
\begin{split}
W_{\rm ext}^{(4)} &= - \int_0^\tau dt \frac{d\lambda_1 (t)}{dt} \int dxdy  P[x,t|y] P[y] \frac{\partial V}{\partial \lambda_1} (x, \lambda_1 (t), \lambda_2 (t)) \\
&= -\frac{1}{2} \int_{k'}^{k} d\lambda_1 \int dxdy P[x,t | y]P[y] (x- \mu_y)^2  =  - \frac{1}{2} \int_{k'}^{k} d\lambda_1 \frac{k_{\rm B}T}{\lambda_1} \\
&= \frac{k_{\rm B}T}{2} \ln \frac{k'}{k} = \frac{k_{\rm B}T}{2} \ln \left( 1+ \frac{S}{N} \right),
\end{split}
\label{work4}
\end{equation}
where we used the fact that the expansion is quasi-static.  Comparing Eqs.~(\ref{work3}) and (\ref{work4}) with mutual information (\ref{Gauss_mutual}),  we obtain
\begin{equation}
W_{\rm ext} := W_{\rm ext}^{(3)}  + W_{\rm ext}^{(4)} = k_{\rm B}T I,
\end{equation}
which achieves the upper bound of inequality~(\ref{second_feedback}).  
%In Ref.~\cite{Abreu}, the optimal feedback protocol that maximizes the extracted work per unit time (i.e., the power) has been discussed.

\subsection{Experimental Demonstration: Feedback-Controlled Ratchet}

We next discuss a recent experiment that realized an information heat engine by using a real-time feedback control on a colloidal particle in water at room temperature~\cite{Toyabe3}.

In the experiment, a colloidal particle with diameter 300 nm is attached to the cover glass, and another particle is attached to the first one [Fig.~7 (a)].  The second particle then moves around the first as a rotating Brownian particle which we observe and control.
An AC electric field is applied with four electrodes, and the particle undergoes an effective potential as illustrated in Fig.~7 (b).  We note that the potential can take two configurations depending on the phases of the electric field. 
Each configuration consists of a spatially-periodic potential and a constant slope.
The slope is created by a constant torque around the circle along which the particle rotates.  This potential is like spiral stairs.  
The depth of the periodic potential is about $3k_{\rm B}T$, and the gradient of the slope per angle $2\pi$ is about $k_{\rm B}T$.  

If the periodic potential without the slope was asymmetric and the two potential configurations were periodically switched, the particle would be transported in one direction as a flashing ratchet~\cite{Vale,Prost,Parrondo2,Reimann,Hanggi2}.  In the present setup, however, the periodic potential without the slope is symmetric and is not switched periodically but switched in a manner that depends on the measured position of the particle via feedback control~\cite{Cao1,Lopez,Touchette3,Parrondo2}.  Such a feedback-controlled ratchet has been experimentally realized~\cite{Lopez}.  In the present experiment, the work and the free energy were measured precisely for quantitatively comparing the experimental results with the theoretical bound~(\ref{second_feedback}). It has been pointed out~\cite{Parrondo2,Bier} that the feedback-controlled ratchet, as well as the flashing ratchet, can be a model of biological molecular motors~\cite{Schliwa}.

The feedback protocol  in the experiment~\cite{Toyabe3} was done as follows [see Fig.~7 (c)].  The position of the particle was probed every 40 ms by a microscope, a camera, and an image analyzer.  Only if the particle was found in the switching region  described by ``S'' in Fig.~7 (c), the potential configuration was switched after a short delay time $\varepsilon$.  By this switching, when the particle reached the hilltop, the potential is inverted, so that the peak of the potential changed into the bottom of the valley, and therefore the particle is transported to the right direction.  Without the switching, the particle would be more likely to go back to the left valley.  This position-dependent switching via feedback control induces the reduction of the entropy in a manner analogous to the feedback control in the Szilard engine.  By performing this protocol many times,  the particle is expected to be transported to the right direction by climbing up the potential slope.
\begin{figure}[htbp]
 \begin{center}
 \includegraphics[width=100mm]{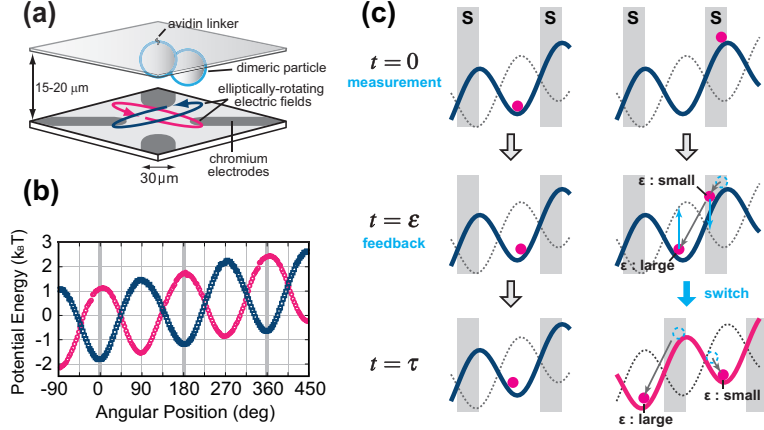}
 \end{center}
 \caption{Experimental setup of the feedback-controlled ratchet (reproduced from Ref.~\cite{Toyabe3} with permission).  (a) A rotating Brownian particle.  (b)  Two possible configurations of the potential that can be switched into each other. (c)  Feedback protocol.  Only if the particle is  found in region ``S,'' the potential is switched. } 
\end{figure}

Figure 8 (a) shows typical trajectories of the particle.  If the feedback delay $\varepsilon$ is sufficiently shorter than the relaxation time of the particle in each well ($\simeq 10$ ms), the particle climbs up the potential.  If the feedback delay is longer than the relaxation time, the feedback does not work and the particle moves down the potential  in agreement with the conventional second law of thermodynamics.  Figure 8 (c) shows the averaged velocity of the particle versus $\varepsilon$, which implies that the shorter the feedback delay is, the faster the average velocity is.

Figure 8 (c) shows the energy balance of this engine.  The shaded region is prohibited by the conventional second law of thermodynamics $\langle  \Delta F - W \rangle \leq 0$, where  $\Delta F$ is the free-energy difference corresponding to the height of the potential, $W$ is the work performed on the particle during the switching, and $\langle \cdots \rangle$ represents the ensemble average over all trajectories.  By using information via feedback, however, the shaded region is indeed achieved if $\varepsilon$ is sufficiently small.  The resource of the excess free-energy gain is thermal fluctuations of the heat bath, which are rectified by feedback control.  This is an experimental realization of an information heat engine.
\begin{figure}[htbp]
 \begin{center}
 \includegraphics[width=80mm]{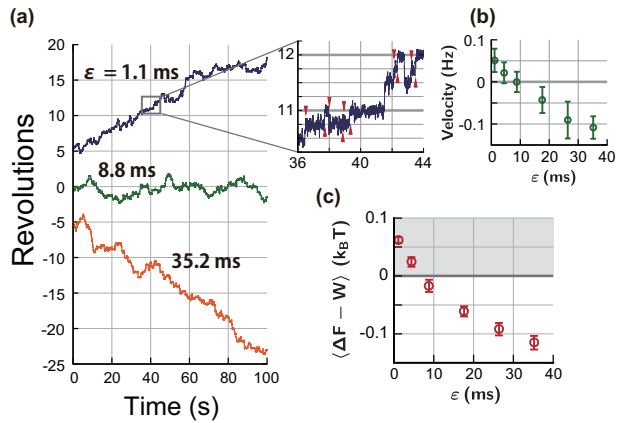}
 \end{center}
 \caption{Experimental results on the feedback-controlled ratchet (reproduced from Ref.~\cite{Toyabe3} with permission).  (a) Typical trajectories of the particle with feedback delays $\varepsilon = 1.1$ ms, $8.8$ ms, and $35.2$ ms.  (b)  The averaged velocity of the particle versus the feedback delay.  (c) Energy balance of feedback control.  The shaded region is prohibited by the conventional second law of thermodynamics, and can only be achieved by feedback control.} 
\end{figure}

For the case of $\varepsilon = 1.1$ ms, $\langle \Delta F - W \rangle  = 0.062 k_{\rm B}T$. On the other hand, the obtained information is given by $I=0.22$, which can be calculated from the histogram of the measurement outcomes by assuming that the measurement is error-free.  By comparing this experimental data with the theoretical bound (\ref{second_feedback}), the efficiency  of this information heat engine is determined to be $
\langle \Delta F - W \rangle / k_{\rm B}TI = 0.062 / 0.22   \simeq 0.28$.
The reason why the efficiency is less than unity is twofold: (1) the switching is not quasi-static but instantaneous, and (2) the obtained information is not utilized if the particle is found outside of the switching region.

We note that various experiments that are analogous to Maxwell's demon have been performed with, for example, a granular gas~\cite{Schlichting,Weele}, supramolecules~\cite{Serreli}, and ultracold atoms~\cite{Raizen}; however, these examples do not explicitly involve the measurement and feedback, as the controlled system and the demon constitute an autonomous system in those experiments.   
Such autonomous versions of Maxwell's demon have also been theoretically studied~\cite{Millonas,Jayannavar,Eggers,Brey,Broeck,Ruschhaupt}.
In contrast, in the experiment of Ref.~\cite{Toyabe3}, the demon (the camera and the computer) is separated from the controlled system (the colloidal particle) as in the case for the Szilard engine.  We also note that an information heat engine similar to that in Ref.~\cite{Toyabe3} has been proposed for an electron pump system in Ref.~\cite{Pekola}.

\subsection{The Carnot Efficiency with Two Heat Baths}

We next consider the case in which there are two heat baths and the process is a thermodynamic cycle.  Without feedback control, the heat efficiency  is bounded by the Carnot bound.  If we perform measurements and feedback control on this system, the extractable work is bounded from above by~\cite{Sagawa-Ueda1}
\begin{equation}
W_{\rm ext} \leq \left( 1 - \frac{T_{\rm L}}{T_{\rm H}} \right) Q_{\rm H} + k_{\rm B}T_{\rm L}I, 
\label{Carnot_feedback}
\end{equation}
where $T_{\rm H}$ and $T_{\rm L}$ are the temperatures of the hot and cold heat baths, respectively, and $Q_{\rm H}$ is the heat that is absorbed by the engine from the hot heat bath.   The proof of inequality (\ref{Carnot_feedback}) will be given in Sec.~5.  The last term on the right-hand side (rhs) of (\ref{Carnot_feedback}) describes the effect of feedback.  We note that the coefficient of the last term is given by the temperature of the cold bath. 

The equality in (\ref{Carnot_feedback}) is achieved in the following example.  We consider a single-particle gas in a box, and quasi-statically control it as in the case of the usual Carnot cycle.  
We then perform the Szilard-engine-type operation consisting of measurement and feedback  on the system while it is in contact with the cold bath.  In this case, we have $k_{\rm B}T_{\rm L}\ln 2$ of excess work, and $Q_{\rm H}$ remains unchanged.  Therefore,  the equality in (\ref{Carnot_feedback}) is achieved with $I = \ln 2$.

We can also achieve the equality in (\ref{Carnot_feedback}) if we perform the Szilard-engine-type operation while the engine is in contact with the hot heat bath.  In this case, we can extract  $k_{\rm B}T_{\rm H}I$ of excess work, and $Q_{\rm H}$ is increased by $k_{\rm B}T_{\rm H}I$.  Therefore, we again obtain the equality in (\ref{Carnot_feedback}) with $I = \ln 2$.

%%%%%%%%%%%%%%%%%%%%%%%%%%%%%%%%%%%%%%%%%%%%%%%%%%%%%%%

\section{Nonequilibrium Equalities with Feedback Control}

Since the late 1990's, a number of universal equalities have been found for nonequilibrium processes~\cite{Jarzynski1,Crooks1,Crooks2,Jarzynski2,Seifert,Kawai,Bustamante,Liphardt,Collin}, and they have been shown to reproduce the second law of thermodynamics and the fluctuation-dissipation theorem.  The fluctuation theorem and the Jarzynski equality are two prime examples of the nonequilibrium equalities.  
In this section, we generalize the nonequilibrium equalities to situations in which a thermodynamic system is subject to measurements and feedback control in line with Refs.~\cite{Sagawa-Ueda3,Horowitz1,Sagawa-Ueda4}. 
As corollaries, we derive inequalities (\ref{second_feedback}) and (\ref{Carnot_feedback}).

\subsection{Preliminaries}

First of all, we review the nonequilibrium equalities without feedback control.
We consider a stochastic thermodynamic system in contact with heat bath(s) with inverse temperatures $\beta_m$ ($m = 1, 2, \cdots$).  Let $x$ be the phase-space point of the system.
The system is controlled through external parameters $\lambda$, which describe, for instance, the volume of a gas or the frequency of an optical tweezer.  Even when the initial state of the system is in thermal equilibrium, the system can be driven far from equilibrium by changing the external parameters.   We consider such a stochastic dynamics of the system from time $0$ to $\tau$.  
The state of the system stochastically evolves as $x(t)$ under a deterministic protocol of the external parameters denoted collectively as $\lambda (t)$.

Let $X_\tau := \{ x(t) \}_{0 \leq t \leq \tau}$ be the trajectory of the phase-space point and $\Lambda_\tau := \{ \lambda(t) \}_{0 \leq t \leq \tau}$ be that of the external parameters.  The heat that is absorbed by the system from heat bath ``$m$''  is a trajectory-dependent quantity, which we write as $Q_m [X_\tau | \Lambda_\tau]$.  The work that is performed on the system is also trajectory-dependent and is denoted as $W [X_\tau | \Lambda_\tau]$.  The first law of thermodynamics is then given by
\begin{equation}
H[x(\tau) | \lambda (\tau )] - H [x(0) | \lambda (0)] = W[X_\tau | \Lambda_\tau] + \sum_m Q_m [X_\tau | \Lambda_\tau], 
\end{equation}
where $H[x | \lambda]$ is the Hamiltonian with external parameters $\lambda$, and the work can be written as
\begin{equation}
W[X_\tau | \Lambda_\tau] = \int_0^\tau \frac{ \partial H}{\partial \lambda} [x(t) | \lambda (t)] \frac{d \lambda (t)}{dt} dt.
\end{equation}

%We next discuss the nonequilibrium equalities and the second law of thermodynamics in this setup.
Let $P [X_\tau | \Lambda_\tau]$ be the probability density of trajectory $X_\tau$ with control protocol $\Lambda_\tau$. It can be decomposed as $P [X_\tau | \Lambda_\tau] = P [X_\tau | x(0),  \Lambda_\tau] P_{\rm f}[x(0)]$, where  $P [X_\tau | x(0),  \Lambda_\tau]$ is the probability density under the condition that the initial state is $x(0)$, and $P_{\rm f}[x(0)]$ is the initial distribution of the forward process.

We next introduce backward processes of the system.
Let $x^\ast$ be the time-reversal of $x$.  For example, if $x = (\bm r, \bm p)$ with $\bm r$ being the position and $\bm p$ being the momentum, then $x^\ast = (\bm r, - \bm p)$.  Similarly, we denote the time-reversal of $\lambda$ as $\lambda^\ast$.  For example, if $\lambda$ is the magnetic field, then $\lambda^\ast = -\lambda$.
 Let $X_\tau^\dagger$ be the time-reversed trajectory of $X_\tau$ defined as $X_\tau^\dagger := \{ x^\ast (\tau - t) \}_{0 \leq t \leq \tau}$.  We also write  the time-reversal of control protocol $\Lambda_\tau$ as $\Lambda^\dagger_\tau := \{ \lambda^\ast (\tau - t) \}_{0 \leq t \leq \tau} $, and write as $P[X^\dagger_\tau | \Lambda^\dagger_\tau]$ the probability density of the time-reversed trajectory with the time-reversed control protocol.  We can decompose $P[X^\dagger_\tau | \Lambda^\dagger_\tau]$ as $P [X_\tau^\dagger | \Lambda_\tau^\dagger] = P [X_\tau^\dagger | x^\dagger (0),  \Lambda_\tau] P_{\rm b}[x^\dagger(0)]$,
where  $P [X_\tau^\dagger | x^\dagger(0),  \Lambda_\tau^\dagger]$ is the probability density under the condition that the initial state of the backward process is $x^\dagger(0)$, and $P_{\rm b}[x^\dagger(0)]$ the initial distribution of the backward process.
We stress that the initial distribution of the backward process $P_{\rm b}[x^\dagger(0)]$ can be set independently of the final distribution of the forward process.  In experiments, we can initialize the system before we start a backward process so that its initial distribution can be chosen independently of the forward process.

The detailed fluctuation theorem (the transient fluctuation theorem) is given by~\cite{Crooks1,Crooks2,Jarzynski2,Seifert}
\begin{equation}
\frac{P [X_\tau^\dagger | x^\dagger (0), \Lambda_\tau^\dagger]}{P [X_\tau | x(0), \Lambda_\tau]} = e^{\sum_m \beta_m Q_m [X_\tau | \Lambda_\tau] }.
\label{fluctuation1}
\end{equation}
Defining the entropy production as
\begin{equation}
\sigma [X_\tau | \Lambda_\tau] := \ln P_{\rm f}[x(0)] - \ln P_{\rm b}[x^\dagger(0)] - \sum_m Q_m[X_\tau | \Lambda_\tau],
\label{entropy_production1}
\end{equation}
we obtain
\begin{equation}
\frac{P [X_\tau^\dagger | \Lambda_\tau^\dagger]}{P [X_\tau | \Lambda_\tau]} = e^{-\sigma [X_\tau | \Lambda_\tau]}.
\label{fluctuation2}
\end{equation}
By taking the ensemble average of Eq.~(\ref{fluctuation2}), we have 
\begin{equation}
\int dX_\tau P [X_\tau | \Lambda_\tau] e^{-\sigma [X_\tau | \Lambda_\tau]} = \int dX_\tau P [X_\tau | \Lambda_\tau]  \frac{P [X_\tau^\dagger | \Lambda_\tau^\dagger]}{P [X_\tau | \Lambda_\tau]}= \int dX_\tau^\dagger P [X_\tau^\dagger | \Lambda_\tau^\dagger]  = 1,
\end{equation}
where we used $dX_\tau = dX_\tau^\dagger$.  Therefore, we obtain the integral fluctuation theorem
\begin{equation}
\langle e^{-\sigma} \rangle = 1.
\label{integral_fluctuation}
\end{equation}
By using the concavity of the exponential function, we find from Eq.~(\ref{integral_fluctuation}) that
\begin{equation}
\langle \sigma \rangle \geq 0,
\label{second1}
\end{equation}
which is an expression of the second law of thermodynamics.  In the following, we discuss the physical meanings of entropy production $\sigma$ for typical situations.
The equality in (\ref{second1}) is achieved if $P [X_\tau^\dagger | \Lambda_\tau^\dagger] = P [X_\tau | \Lambda_\tau]$ holds for any $X_\tau$, which implies the reversibility of the process.

We first consider  isothermal processes.  In this case, we choose the initial distributions of the forward and backward processes as 
\begin{equation}
\begin{split}
P_{\rm f} [x(0)] &= \exp \left( \beta ( F[\lambda (0)] - H [x(0) | \lambda (0)]) \right), \\
P_{\rm b} [x^\dagger(0)] &= \exp \left( \beta ( F[\lambda^\dagger (0)] - H [x^\dagger (0) | \lambda^\dagger (0)]) \right), 
\end{split}
\label{initial}
\end{equation}
where $F[\lambda] := - k_{\rm B}T \ln \int dx e^{-\beta H[x | \lambda]}$ is the free energy of the system.
We assume that the Hamiltonian has the time-reversal symmetry
\begin{equation}
H[x | \lambda] = H[x^\ast | \lambda^\ast],
\end{equation}
and therefore the canonical distribution satisfies
\begin{equation}
P_{\rm b} [x^\dagger(0)] = \exp \left( \beta ( F[\lambda (\tau)] - H [x (\tau) | \lambda (\tau)]) \right).
\end{equation}
The entropy production then reduces to
\begin{equation}
\sigma [X_\tau] = \beta (W [X_\tau] - \Delta F),
\label{entropy_production2}
\end{equation}
where $\Delta F := F[\lambda (\tau)] - F[\lambda (0)]$.
Thus, the integral fluctuation theorem~(\ref{integral_fluctuation})  leads to the Jarzynski equality~\cite{Jarzynski1}
\begin{equation}
\langle e^{-\beta W} \rangle = e^{-\beta \Delta F},
\label{Jarzynski}
\end{equation}
and inequality~(\ref{second1}) gives the second law of thermodynamics
\begin{equation}
\langle W \rangle \geq \Delta F,
\label{second2}
\end{equation}
where $W_{\rm ext} := - \langle W \rangle$ is the work that is extracted from the system.

We next consider a case with multi-heat baths, and  assume that the initial distributions of the forward and the backward processes are given by the canonical distributions as in~(\ref{initial}) with a reference inverse temperature $\beta$. 
In practice, $\beta$ can be taken as one of $\beta_m$'s, which can be realized if the system is initially attached only to that particular heat bath.
We then have
\begin{equation}
\sigma [X_\tau] = \beta (\Delta E [X_\tau] -\Delta F) - \sum_m \beta_m Q_m [X_\tau],
\label{entropy_production3}
\end{equation}
where $\Delta E[X_\tau] := H [x (\tau) | \lambda (\tau)] - H [x (0) | \lambda (0)]$ is the difference of the internal energy of the system.
Inequality~(\ref{second1}) leads to
\begin{equation}
\sum_m \beta_m \langle Q_m  \rangle \leq \beta ( \langle \Delta E \rangle - \Delta F ).
\label{second3}
\end{equation}
In particular, if the process is a cycle such that $\Delta F = 0$ and $\langle \Delta E \rangle = 0$ hold, inequality~(\ref{second3}) reduces to
\begin{equation}
\sum_m \beta_m \langle Q_m  \rangle \leq 0,
\label{Clausius}
\end{equation}
which is the Clausius inequality.  
If there are two heat baths with temperatures $T_{\rm H}$ and $T_{\rm L}$, (\ref{Clausius}) gives the Carnot bound
\begin{equation}
- \langle W \rangle \leq \left( 1 - \frac{T_{\rm L}}{T_{\rm H}} \right) \langle Q_{\rm H} \rangle,
\end{equation}
where $\langle Q_{\rm H} \rangle$ is the average of the heat that is absorbed by the engine from the hot heat bath.

\subsection{Measurement and Feedback}

We now formulate measurements and feedback on the thermodynamic system~\cite{Sagawa-Ueda3,Horowitz1,Sagawa-Ueda4}.
We perform measurements at time $t_k$ ($k=1, 2, \cdots, M$) with $0 \leq t_1 < t_2 < \cdots < t_M < \tau$.  Let $y (t_k)$ be the measurement outcome at time $t_k$.  For simplicity, we assume that the measurement is instantaneous; the measurement error of $y (t_k)$ can be characterized only by the conditional probability $P_{\rm c}[y( t_k) | x(t_k)]$, which implies that only $y (t_k)$ has the information about $x(t_k)$. (Note, however, that this assumption can be relaxed~\cite{Sagawa-Ueda4}.)
We write the sequence of the measurement outcomes as $Y_\tau := (y(t_1), y(t_2), \cdots, y(t_M))$, and write
\begin{equation}
P_{\rm c}[Y_\tau | X_\tau] := \prod_k P_{\rm c}[y (t_k) | x(t_k)].
\label{measurement_probability}
\end{equation}
We then introduce the following quantity:
\begin{equation}
I_{\rm c} [X_\tau: Y_\tau] := \ln \frac{P_{\rm c}[Y_\tau | X_\tau]}{P[Y_\tau]},
\label{stochastic_mutual}
\end{equation}
which  can be interpreted as a stochastic version of the mutual information. The ensemble average of Eq.~(\ref{stochastic_mutual}) gives the mutual information obtained by the measurements as
\begin{equation}
\langle I_{\rm c} \rangle = \int dX_\tau dY_\tau P[X_\tau, Y_\tau] \ln \frac{P_{\rm c}[Y_\tau | X_\tau]}{P[Y_\tau]}.
\label{average_mutual}
\end{equation}
We note that $\langle I_{\rm c} \rangle$ describes the correlation between $X_\tau$ and $Y_\tau$ that is induced only by measurements, and not by feedback control.  The suffix ``c'' represents this property of $I_{\rm c}$.  We then identify $\langle I_{\rm c} \rangle$ with $I$ in the foregoing arguments.  See Ref.~\cite{Sagawa-Ueda4} for details. 
We note that $\langle I_{\rm c} \rangle$ has been discussed and referred to as the transfer entropy in Ref.~\cite{Schreiber}.
The following equality also holds by definition: 
\begin{equation}
\langle e^{- I_{\rm c}} \rangle = 1.
\label{fluctuation_information}
\end{equation}

We next consider feedback control by using the obtained outcomes.   The control protocol after time $t_k$ can depend on the outcome $y (t_k)$ in the presence of feedback control.  We write this dependence as $\Lambda_\tau (Y_\tau)$.
We can  show that the joint probability of $(X_\tau, Y_\tau)$ is given by
\begin{equation}
P[X_\tau, Y_\tau ] = P_{\rm c}[Y_\tau | X_\tau] P[X_\tau | \Lambda (Y_\tau)],
\label{joint_probability}
\end{equation}
which satisfies the normalization condition $\int dX_\tau dY_\tau P[X_\tau, Y_\tau] = 1$.  Equality~(\ref{joint_probability}) is proved in Appendix A (see also Ref.~\cite{Sagawa-Ueda4}).  
The probability of obtaining outcome $Y_\tau$ in the forward process is  given by the marginal distribution as $P[Y_\tau] = \int dX_\tau P[X_\tau, Y_\tau ] $,
and the conditional probability is given by $P[X_\tau | Y_\tau] = P[X_\tau, Y_\tau]/P[Y_\tau]$.
In the presence of measurement and feedback, the ensemble average is taken over all trajectories and all outcomes; for an arbitrary stochastic quantity $A[X_\tau, Y_\tau]$, its ensemble average is given by
\begin{equation}
\langle A \rangle = \int dX_\tau dY_\tau P[X_\tau, Y_\tau] A[X_\tau, Y_\tau].
\label{average}
\end{equation}

The detailed fluctuation theorem for a given $Y_\tau$ can be written as
\begin{equation}
\frac{P [X_\tau^\dagger | \Lambda_\tau (Y_\tau)^\dagger]}{P [X_\tau | \Lambda_\tau (Y_\tau)]} = e^{-\sigma [X_\tau | \Lambda_\tau (Y_\tau)]}.
\label{fluctuation3}
\end{equation}
We note that Eq.~(\ref{fluctuation3}) is valid in the presence of feedback control, because the detailed fluctuation theorem is satisfied once a control protocol is fixed.
Equality~(\ref{fluctuation3}) provides the basis for the derivations of the formulas in the following section.

\subsection{Nonequilibrium Equalities  with Mutual Information}

In this subsection, we generalize the nonequilibrium equalities by incorporating the mutual information.
First of all, from Eq.~(\ref{stochastic_mutual}), we have
\begin{equation}
\frac{P[Y_\tau]}{P_{\rm c}[Y_\tau | X_\tau]} = e^{-I_{\rm c}}.
\label{stochastic_mutual2}
\end{equation}
By multiplying the both-hand sides of this equality by those of Eq.~(\ref{fluctuation3}), we obtain
\begin{equation}
\frac{P [X_\tau^\dagger | \Lambda_\tau (Y_\tau)^\dagger] P[Y_\tau ]}{P[X_\tau, Y_\tau]} = e^{-\sigma [X_\tau | \Lambda_\tau (Y_\tau)] - I_{\rm c}[X_\tau : Y_\tau]}.
\label{fluctuation_feedback1}
\end{equation}
To measure $P [X_\tau^\dagger | \Lambda_\tau (Y_\tau)^\dagger] P[Y_\tau ]$, we follow the  backward process corresponding to each forward outcome and count the occurrences of the time-reversed trajectories.  By taking the ensemble average of the both-hand sides of Eq.~(\ref{fluctuation_feedback1}) with formula~(\ref{average}), we obtain a generalized integral fluctuation theorem with feedback control:
\begin{equation}
\langle e^{-\sigma - I_{\rm c}} \rangle = 1.
\label{integral_fluctuation_feedback1}
\end{equation}
Using the concavity of the exponential function, Eq.~(\ref{integral_fluctuation_feedback1}) leads to 
\begin{equation}
\langle \sigma \rangle \geq - \langle I_{\rm c} \rangle.
\label{second_feedback1}
\end{equation}
Inequality~(\ref{second_feedback1}) is a generalized second law of thermodynamics, which states that the entropy production can be decreased by feedback control, and that the lower bound of the entropy production is given by the mutual information $\langle I_{\rm c} \rangle$.  As shown below, inequalities~(\ref{second_feedback}) and (\ref{Carnot_feedback}) in Sec.~4 are special cases of inequality~(\ref{second_feedback1}).
The equality in (\ref{second_feedback1}) is achieved if $P [X_\tau^\dagger | \Lambda_\tau (Y_\tau)^\dagger] P[Y_\tau ] = P[X_\tau, Y_\tau]$ holds for any $X_\tau$ and $Y_\tau$, which implies the reversibility with feedback control as discussed in Ref.~\cite{Horowitz2}.

The generalized integral fluctuation theorem of the form (\ref{integral_fluctuation_feedback1}) was first shown in Ref.~\cite{Sagawa-Ueda3} for a single measurement, and Eq.~(\ref{fluctuation_feedback1}) was obtained in Ref.~\cite{Horowitz1,Sagawa-Ueda4} for multiple measurements.  These results has also been generalized to the optimal control process with continuous measurement and the Kalman filter in Ref.~\cite{Suzuki}. 

A generalized fluctuation theorem was also obtained in Ref.~\cite{Kim}, which is similar to Eq.~(\ref{integral_fluctuation_feedback1}).  In Ref.~\cite{Kim}, feedback control is performed based on information about the continuously-monitored velocity of a Langevin system.  The result of  Ref.~\cite{Kim} includes an quantity that describes the decrease in the entropy by continuous feedback control, instead of the mutual information obtained by the continuous measurement.  

We consider isothermal processes with a single heat bath, in which the entropy production is given by Eq.~(\ref{entropy_production2}).  Equality~(\ref{integral_fluctuation_feedback1})  then reduces to a generalized Jarzynski equality
\begin{equation}
\langle e^{\beta (\Delta F - W) - I_{\rm c}} \rangle = 1,
\label{Jarzynski_feedback1}
\end{equation}
and inequality~(\ref{second_feedback1}) reduces to 
\begin{equation}
\langle \Delta F - W \rangle \geq k_{\rm B}T \langle I_{\rm c} \rangle,
\label{second_feedback2}
\end{equation}
which implies inequality (\ref{second_feedback}) with identifications  $W_{\rm ext} = - \langle W \rangle$, $\Delta F = \langle \Delta F \rangle$, and $I = \langle I_{\rm c} \rangle$.

We next consider the cases in which there are two heat baths and the process is a cycle, in which the entropy production is given by the ensemble average of Eq.~(\ref{entropy_production3}) with $\langle \Delta E \rangle = \langle \Delta F \rangle = 0$.
The generalized second law~(\ref{second_feedback1}) then leads to
\begin{equation}
\beta_{\rm H} \langle Q_{\rm H} \rangle + \beta_{\rm L} \langle Q_{\rm L} \rangle \leq \langle I_{\rm c} \rangle,
\end{equation}
which can be rewritten as
\begin{equation}
- \langle W \rangle \leq \left( 1 - \frac{T_{\rm L}}{T_{\rm H}} \right) \langle Q_{\rm H} \rangle + k_{\rm B}T_{\rm L} \langle I_{\rm c} \rangle.
\label{second_feedback3}
\end{equation}
By identifying $ Q_{\rm H} = \langle Q_{\rm H} \rangle$, inequality~(\ref{second_feedback3}) implies inequality (\ref{Carnot_feedback}).

\subsection{Nonequilibrium Equalities with Efficacy Parameter}

In this subsection, we discuss another generalization of nonequilibrium equalities.  We define the time-reversal of outcomes $Y_\tau$ as $Y_\tau^\dagger := (y(\tau - t_M)^\ast, \cdots, y(\tau - t_2)^\ast, y(\tau - t_1)^\ast)$, where $y^\ast$ is the time-reversal of $y$, and  introduce the probability that we obtain outcome $Y_\tau^\dagger$ with control protocol $\Lambda_\tau (Y_\tau)^\dagger$, which is given by
\begin{equation}
P [Y_\tau^\dagger | \Lambda_\tau (Y_\tau)^\dagger ] = \int dX_\tau^\dagger P_{\rm c}[Y_\tau^\dagger | X_\tau^\dagger ] P[X_\tau^\dagger | \Lambda_\tau (Y_\tau)^\dagger].
\label{reverse_probability}
\end{equation}
We stress that no feedback control is performed in the backward processes.
We then assume that the measurement error has the time-reversal symmetry
\begin{equation}
P_{\rm c}[Y_\tau^\dagger | X_\tau^\dagger ]  = P_{\rm c}[Y_\tau | X_\tau ].
\label{assumption_measurement}
\end{equation}
This assumption is satisfied if $P_{\rm c}[y(t_k) | x(t_k)] = P_{\rm c}[y(\tau - t_k)^\ast | x(\tau - t_k)^\ast]$ holds for $k = 1, 2, \cdots, M$.
By using Eq.~(\ref{reverse_probability}) and assumption (\ref{assumption_measurement}), we can show that
\begin{equation}
\frac{P [Y_\tau^\dagger |   \Lambda_\tau (Y_\tau)^\dagger] }{P[Y_\tau]} = \langle e^{- \sigma} \rangle_{Y_\tau},
\label{KPB}
\end{equation}
where $\langle \cdots \rangle_{Y_\tau}$ denotes the conditional average with condition $Y_\tau$ such that
\begin{equation}
\langle e^{- \sigma} \rangle_{Y_\tau} := \int dX_\tau e^{-\sigma [X_\tau | \Lambda_\tau (Y_\tau)]} P[X_\tau | Y_\tau].
\end{equation}
Equality~(\ref{KPB}) has been shown for Hamiltonian systems~\cite{Kawai} and stochastic systems~\cite{Sagawa-Ueda3,Sagawa-Ueda4}.
By noting that
\begin{equation}
\langle e^{- \sigma } \rangle = \int dY_\tau P[Y_\tau] \langle e^{-\sigma} \rangle_{Y_\tau},
\end{equation}
we obtain yet another generalization of the integral fluctuation theorem~\cite{Sagawa-Ueda3,Sagawa-Ueda4}
\begin{equation}
\langle e^{- \sigma } \rangle = \gamma,
\label{integral_fluctuation_feedback2}
\end{equation}
where
\begin{equation}
\gamma = \int dY_\tau P [Y_\tau^\dagger | \Lambda_\tau (Y_\tau)^\dagger ]
\label{gamma}
\end{equation}
is the sum of the probabilities that we obtain the time-reversed outcomes with a time-reversed protocol.  
For the cases of isothermal processes, Eq.~(\ref{integral_fluctuation_feedback2}) reduces to
\begin{equation}
\langle e^{\beta (\Delta F - W)} \rangle = \gamma.
\label{Jarzynski_feedback2}
\end{equation}

We note that $\gamma$ characterizes the efficacy of feedback control.  The more efficient the feedback protocol is, the larger the amount of $\gamma$ is.
Without feedback control, $P [Y_\tau^\dagger | \Lambda_\tau^\dagger ]$ reduces to a single unconditional probability distribution, and we therefore obtain
\begin{equation}
\gamma = \int dY_\tau P [Y_\tau^\dagger | \Lambda_\tau^\dagger ] = 1,
\end{equation}
which reproduces the integral fluctuation theorem (\ref{integral_fluctuation}) without feedback.
We note that the maximum value of $\gamma$ is the number of the possible outcomes of $Y_\tau$.

We illustrate the efficacy parameter $\gamma$ for the case of the Szilard engine that is described in Sec.~2.  The backward control protocol of the Szilard engine is as follows (see also Fig.~9)~\cite{Sagawa-Ueda3}.
\begin{figure}[htbp]
 \begin{center}
 \includegraphics[width=70mm]{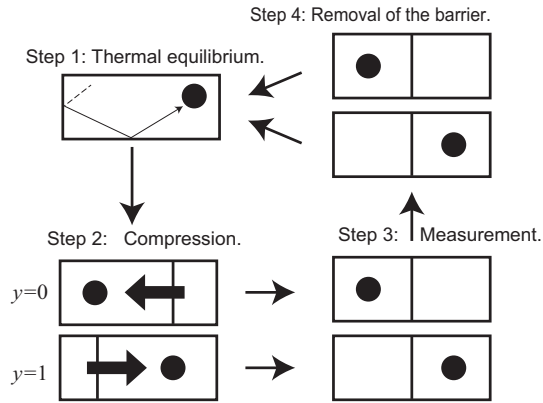}
 \end{center}
 \caption{Backward processes of the Szilard engine.  See the text for details.} 
\end{figure}

\

\textit{Step 1: Initial state.}
The single-particle gas is initially in thermal equilibrium.

\

\textit{Step 2: Compression of the box.} In accordance with the measurement outcome in the forward process, which is ``$0$'' ($=$ ``left'') or ``$1$'' ($=$ ``right''),  we quasi-statically compress the box by moving the wall in the box to the center.  By this compression, the volume of the box becomes half.

\

\textit{Step 3: Measurement.} We  measure the position of the particle to find which box the particle is in.  The outcome of this backward measurement is ``$0$'' ($=$ ``left'') or ``$1$'' ($=$ ``right'') with unit probability corresponding to forward outcome ``$0$'' or ``$1$,'' respectively.

\

\textit{Step 4:}  We remove the barrier at the center of the box, and the engine returns to the initial state by a free expansion.

\

In these backward processes, the measurement outcomes in \textit{step 2} satisfy $P[0 | \Lambda_\tau (0)^\dagger] = 1$ and $P[1 | \Lambda_\tau (1)^\dagger] = 1$,
and therefore we obtain $\gamma = P[0 | \Lambda_\tau (0)^\dagger] + P[1 | \Lambda_\tau (1)^\dagger] = 2$,
which gives the maximum value of $\gamma$ for  situations in which the number of possible outcomes is two.
On the other hand,  since $W = - k_{\rm B}T \ln 2$ and $\Delta F = 0$  in the absence of  fluctuations,  the generalized Jarzynski equality (\ref{Jarzynski_feedback2}) is satisfied as $\langle e^{\beta (\Delta F - W)} \rangle = 2 = \gamma$.

The generalized Jarzynski equality (\ref{Jarzynski_feedback2}) has been experimentally verified in the experiment described in Sec.~4.4 by measuring $\Delta F - W$ and $\gamma$ separately in the forward and backward experiments, respectively~\cite{Toyabe3}.   Equalities~(\ref{Jarzynski_feedback1}) and (\ref{Jarzynski_feedback2}) have been obtained in Hamiltonian systems~\cite{Sagawa}.  
Equality~(\ref{Jarzynski_feedback2}) has also been generalized to quantum systems~\cite{Morikuni,Lahiri}.

While Eq.~(\ref{integral_fluctuation_feedback1}) only includes the obtained mutual information and does not describe how we utilize the information via feedback,  Eq.~(\ref{integral_fluctuation_feedback2}) includes the term of feedback efficacy that depends on the feedback protocol.  To quantitatively discuss the relationship between mutual information $I_{\rm c}$ and efficacy parameter $\gamma$, we define $C[A] := - \ln \langle e^{-A} \rangle$.
By noting Eq.~(\ref{fluctuation_information}), we obtain
\begin{equation}
C[\sigma ] + C[I_{\rm c}] - C[\sigma + I_{\rm c}] = - \ln \gamma.
\label{I_gamma1}
\end{equation}
If the joint distribution of $\sigma$ and $I_{\rm c}$ is Gaussian,  Eq.~(\ref{I_gamma1}) reduces to
\begin{equation}
\langle \sigma I_{\rm c} \rangle - \langle \sigma \rangle \langle I_{\rm c} \rangle  = -\ln \gamma.
\label{I_gamma2}
\end{equation}
Equalities~(\ref{I_gamma1}) and (\ref{I_gamma2}) imply that, the more efficiently we use the obtained information to decrease the entropy production by feedback control, the larger $\gamma$ is.  In fact, if $\gamma$ is large, the left-hand sides of Eqs.~(\ref{I_gamma1}) and (\ref{I_gamma2}) are both small, which means that the obtained information $I_{\rm c}$ has a large  negative correlation with $\sigma$.  Without feedback control, $\gamma = 1$ holds and therefore $I_{\rm c}$ is not correlated with $\sigma$.  In this sense, $\gamma$ characterizes the efficacy of feedback control.

%%%%%%%%%%%%%%%%%%%%%%%%%%%%%%%%%%%%%%%%%%%%%%%%%%%%%%%

\section{Thermodynamic Energy Cost for Measurement and Information Erasure}

So far, we have discussed the energy balance of information heat engines controlled by the demon.  In this section, we discuss the energy cost that is needed for the demon itself, which has been a  subject of active discussion~\cite{Demon,Maruyama,Maroney0,Brillouin,Landauer,Bennett,Zurek2,Zurek3,Shizume,Goto,Piechocinska,Bennett2,Allahverdyan3,Horhammer,Barkeshli,Norton,Maroney,Turgut,Sagawa-Ueda2}.

Suppose that the demon has a memory that can store the outcome obtained by a measurement.  If the outcome is binary, the memory can be modeled by a system with a binary potential (see Fig.~10).   Before the measurement, the memory is in the initial standard state $0$.  The memory then interacts with a measured system such as the Szilard engine, and stores the measurement outcome.  Figure 10 illustrates a case with a binary outcome.  Let $p_k$ be the probability of obtaining  outcome $k$.  After the measurement, the memory is detached from the measured system and returns to the initial standard state, which is the erasure of the obtained information.   The central question in this section is how much work is needed for the demon during the measurement and the information erasure.
\begin{figure}[htbp]
 \begin{center}
 \includegraphics[width=88mm]{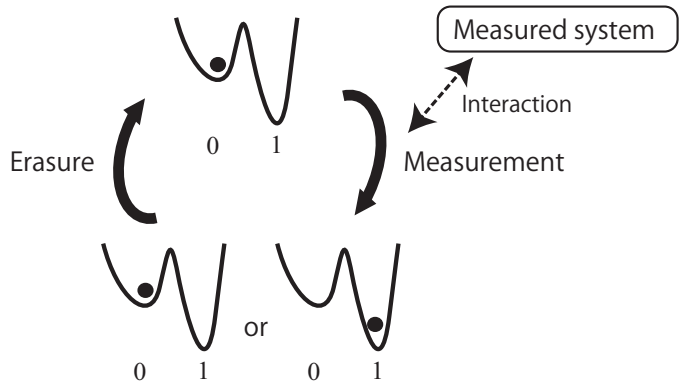}
 \end{center}
 \caption{A schematic of the measurement and erasure of information for the case of an asymmetric binary memory.  While the memory is in the standard state with unit probability  before the measurement, the memory stores the measurement  outcome in accordance with the state of the measured system.  The measurement and erasure processes are time-reversal with each other except for the fact that the memory establishes a correlation with the measured system during the measurement process. } 
\end{figure}

Let $F_k^{\rm M}$ be the free energy of the memory under the condition that the outcome is ``$k$.''  During the measurement process, the free energy of the memory is changed on average by $\Delta F^{\rm M} := \sum_k p_k F_k^{\rm M} - F_0^{\rm M}$, where $F_0^{\rm M}$ is the free energy of the initial standard state.  If $F_k^{\rm M}$'s are the same for all $k$'s including $k=0$ (i.e., the memory's potential is symmetric), $\Delta F^{\rm M} = 0$ holds for every $\{ p_k \}$.  It has been shown~\cite{Sagawa-Ueda2} that  the averaged work $W_{\rm meas}^{\rm M}$ that is performed on the memory during the measurement is bounded as
\begin{equation}
W_{\rm meas}^{\rm M} \geq \Delta F^{\rm M} - k_{\rm B}T H + k_{\rm B}T I,
\label{measurement}
\end{equation}
where $H := - \sum_k p_k \ln p_k$ is the Shannon information of the outcomes and $I$ is the mutual information obtained by the measurement.  For the special case with $\Delta F^{\rm M} = 0$ and $H = I$, the rhs of inequality (\ref{measurement}) reduces to zero.

On the other hand, during the information erasure, the change of the free energy of the memory is given by $- \Delta F^{\rm M}$.  The averaged  work $W_{\rm eras}^{\rm M}$ that is needed for the erasure process is bounded as~\cite{Sagawa-Ueda2}
\begin{equation}
W_{\rm eras}^{\rm M} \geq - \Delta F^{\rm M} + k_{\rm B}T H.
\label{erasure}
\end{equation}
If $\Delta F^{\rm M}$ vanishes, inequality (\ref{erasure}) reduces to 
\begin{equation}
W_{\rm eras}^{\rm M} \geq k_{\rm B}T H,
\label{Landauer}
\end{equation}
which is known as the Landauer principle~\cite{Landauer}. The additional term $-\Delta F^{\rm M}$ on the rhs of (\ref{erasure}) arises from the asymmetry of the memory.
By summing up inequalities (\ref{measurement}) and (\ref{erasure}), we obtain the fundamental inequality
\begin{equation}
W_{\rm meas}^{\rm M} + W_{\rm eras}^{\rm M} \geq k_{\rm B}T I,
\label{total}
\end{equation}
which implies that the work needed for the demon is only bounded by the mutual information if we take into account both the measurement and erasure processes.

We stress that, while inequality (\ref{erasure}) is a generalized Landauer principle for the information erasure, inequality (\ref{total}) is completely different from the Landauer principle.  In fact, while the lower bound of the Landauer principle is given by the Shannon information that characterizes the randomness of the measurement outcomes, the lower bound of (\ref{total}) is given by the mutual information that characterizes the correlation between the measured system and the measurement outcome.
Moreover,  both terms on the rhs of (\ref{erasure}) is exactly canceled by the first and second terms on the rhs of (\ref{measurement}).  The reason for the cancellation lies in the fact that the dynamics of the memory during the erasure process is the time-reversal of the measurement process, except for the fact that the memory interacts with the measured system and establishes a correlation (or equivalently, gains information) only in the measurement process (see also Fig.~10).  The additional cost for the establishment of the correlation is given by the last term on the rhs of (\ref{measurement}), which also appears in the rhs of (\ref{total}).  Therefore, the mutual information term in inequality (\ref{total})  is induced by the measurement process.

Historically, there has been a lot of discussions~\cite{Demon,Maruyama,Maroney0} as to what compensates for the additional work of  $k_{\rm B}T \ln 2$ which can be extracted from the Szilard engine.
Szilard considered that an entropic cost must be needed for the measurement process~\cite{Szilard}.
L.~Brillouin~\cite{Brillouin} argued that we need the work greater than $k_{\rm B}T \ln 2$ for the measurement process, based on a specific model of measurement.  Later, by explicitly constructing a model of the memory does not require any work for the measurement, C. H. Bennett argued that, based on the Landauer principle~(\ref{Landauer}), we always need the work of at least $k_{\rm B}T \ln 2$ for the erasure process~\cite{Bennett,Bennett2}.  The key observation here is that the erasure process is logically irreversible while the measurement process can be logically reversible.  In fact, if we assume that the Shannon information of the measurement outcome equals the thermodynamic entropy of the memory, the logically irreversible erasure should be accompanied by a reduction in  the thermodynamic entropy of the memory, which implies that $k_{\rm B}T \ln 2$ of heat should be transfered to the heat bath and, therefore, the same amount of the work is needed.

However, the argument by Landauer and Bennett is valid only for symmetric memories with $\Delta F^{\rm M} = 0$.  
As discussed in Refs.~\cite{Barkeshli,Norton,Maroney,Turgut,Sagawa-Ueda2}, the Shannon information does not equal the thermodynamic entropy of the memory in general.  If the memory is asymmetric as illustrated in Fig.~10, the lower bound of the energy cost needed for the information erasure is not given by (\ref{Landauer}), and the Landauer principle needs to be generalized to inequality~(\ref{erasure}) for asymmetric memories.   We note that the Landauer principle can also be violated for symmetric memories in the quantum regime due to the initial correlation between the memory and the heat bath~\cite{Allahverdyan3,Horhammer}.
A more detailed historical review about the Landauer principle is given in Ref.~\cite{Maroney0}.

As a consequence, the lower bound of the individual energy cost for measurement or erasure processes can be made arbitrarily small for asymmetric memories, while their sum~(\ref{total}) is  bounded from below by $k_{\rm B}TI$  that originates from the measurement process.  
The total work given in the left-hand-side of (\ref{total}) then compensates for $k_{\rm B}T I$ of  additional work in (\ref{second_feedback}) that is extracted from an information heat engine by the demon.  This compensation  confirms the consistency between the demon and the second law of thermodynamics; we cannot extract any positive amount of work by a cycle from the total system consisting of the engine and the memory of the demon.

Nevertheless, feedback control is still useful for manipulating small thermodynamic systems.  In fact, as discussed in Sec.~2, feedback control enables us to increase the engine's free energy  without injecting energy to the engine directly. In other words, the work (\ref{total}) needed for the demon is not necessarily transfered to the engine, which can be energetically separated from the demon.  Therefore, by using information heat engines, we can control thermodynamic systems beyond the energy balance that is imposed by the conventional thermodynamics.

%%%%%%%%%%%%%%%%%%%%%%%%%%%%%%%%%%%%%%%%%%%%%%%%%%%%%%%%%%

\section{Conclusions}

In this chapter, we have discussed a generalized thermodynamics that can be applied to feedback-controlled systems which we call information heat engines.  The Szilard engine described in Sec.~2 is the simplest model of information heat engines.  Based on the information theory reviewed in Sec.~3, we have formulated a generalized second law involving the term of the mutual information in Sec.~4.  The generalized second law gives an upper bound of the work that can be extracted from a heat bath with the assistance of feedback control. We also discussed some typical examples of information heat engines including a recent experimental result~\cite{Toyabe3}.  In Sec.~5, we discussed nonequilibrium equalities with feedback control, and derived the generalized second law discussed in Sec.~4.  We also discussed the energy cost that is needed for the measurement and the information erasure in Sec.~6.

Inequalities (\ref{second_feedback}), (\ref{measurement}), (\ref{erasure}), and (\ref{total}) are the generalizations of the second law of thermodynamics, giving the fundamental bounds of the work needed for information processing.  In fact, if we set the information contents to be zero (i.e., $I=H=0$) in these inequalities, all of them reduce to the conventional second law of thermodynamics.  In this sense, these inequalities constitute the second law of ``information thermodynamics,'' which is a generalized thermodynamics for information processing.

While the studies of information and thermodynamics have a long history, recent developments of nonequilibrium physics and nanotechnologies have shed new light on classic problems from the modern point of view.  Thermodynamics of information processing will open a fruitful research arena that enables us to quantitatively analyze the energy costs of the feedback control and information processing in small thermodynamic systems.  Possible applications of this new research field include designing designing and controlling nanomachines~\cite{Kay} and nanodevices.

%%%%%%%%%%%%%%%%%%%%%%%%%%%%%%%%%%%%%%%%%%%%%%%%%%%%%%%%%%%%%%%%%%

\appendix

\section{Proof of Eq.~(\ref{joint_probability})}

In this appendix, we prove Eq.~(\ref{joint_probability}).
We introduce notations $X_{t_{k-1} < t \leq t_k} := \{ x(t) \}_{t_{k-1} < t \leq t_k}$, $X_{t_k} := \{ x (t) \}_{0 \leq t \leq t_k}$, $\Lambda_{t_k} := \{ \lambda (t) \}_{0 \leq t \leq t_k}$, and  $Y_{t_k} := (y(t_1), \cdots, y(t_k))$.
The joint probability of $(X_\tau, Y_\tau)$ is given by
\begin{equation}
\begin{split}
P[X_\tau, Y_\tau ] &= P[X_{t_M < t \leq \tau} | X_{t_M}, \Lambda_\tau (Y_\tau)] \\
&\cdot \prod_{k=1}^M P[y(t_k)|x(t_k)] P[X_{t_{k-1} < t \leq t_k} | X_{t_{k-1}}, \Lambda_{t_k} (Y_{t_{k-1}})] \cdot P_{\rm f}[x(0)],
\end{split}
\label{joint_probability0}
\end{equation}
where  we set $t_0 := 0$. 
We note that $\Lambda_{t_k}$ depends only on $Y_{t_{k-1}}$ due to the causality.
We also note that 
\begin{equation}
P[X_\tau | \Lambda (Y_\tau)] = P[X_{t_k < t \leq \tau} | X_{t_k}, \Lambda_\tau (Y_\tau)] \prod_{k=1}^M P[X_{t_{k-1} < t \leq t_k} | X_{t_{k-1}}, \Lambda_{t_k} (Y_{t_{k-1}})] P_{\rm f}[x(0)].
\label{appendix2}
\end{equation}
By combining Eqs.~(\ref{measurement_probability}), (\ref{joint_probability0}), and (\ref{appendix2}), we obtain Eq.~(\ref{joint_probability}).
We can confirm that the joint probability satisfies the normalization condition $\int dX_\tau dY_\tau P[X_\tau, Y_\tau] = 1$ by integrating Eq.~(\ref{joint_probability0}) in the order of $X_{t_M < t \leq \tau} \to y(t_M)$ $\to X_{t_{M-1} < t \leq t_M} \to y(t_{M-1}) \to \cdots \to y(t_1) \to X_{0 < t \leq t_1} \to  x(0)$ due to the causality.

\paragraph{Acknowledgment.}  The authors are grateful to Shoichi Toyabe, Eiro Muneyuki, and Masaki Sano for providing us the experimental data discussed in Sec.~4.4.
This work was supported by KAKENHI 22340114, a Grant-in-Aid for Scientific Research on Innovation Areas "Topological Quantum Phenomena" (KAKENHI 22103005), a Global COE Program "the Physical Sciences Frontier", the Photon Frontier Network Program, and the Grant-in-Aid for Research Activity Start-up (KAKENHI 11025807), from MEXT of Japan.


\begin{thebibliography}{150}

\bibitem{Maxwell} J. C. Maxwell, ``\textit{Theory of Heat},'' (Appleton, London, 1871).
\bibitem{Demon}  \textit{``Maxwell's demon 2: Entropy, Classical and Quantum Information, Computing''}, H. S. Leff and A. F. Rex (eds.), (Princeton University Press, New Jersey, 2003).
\bibitem{Maruyama}  K. Maruyama, F. Nori, and V. Vedral, Rev. Mod. Phys. \textbf{81}, 1 (2009).
\bibitem{Maroney0} O. J. E. Maroney, ``Information Processing and Thermodynamic Entropy'', The Stanford Encyclopedia of Philosophy (Fall 2009 Edition), Edward N. Zalta (ed.).


\bibitem{Doyle} J. C. Doyle, B. A. Francis, and A. R. Tannenbaum, ``\textit{Feedback Control Theory},'' (Macmillan, New York, 1992).
\bibitem{Astrom} K. J. \r{A}strom and R. M. Murray, ``\textit{Feedback Systems: An Introduction for Scientists and Engineers},'' ( Princeton University Press, 2008).

\bibitem{Lloyd1} S. Lloyd, Phys. Rev. A \textbf{39}, 5378 (1989).
\bibitem{Caves} C. M. Caves, Phys. Rev. Lett. \textbf{64}, 2111 (1990).
\bibitem{Lloyd2} S. Lloyd, Phys. Rev. A \textbf{56}, 3374 (1997).
\bibitem{Nielsen} M. A. Nielsen, C. M. Caves, B. Schumacher, and H. Barnum, Proc. R. Soc. London A \textbf{454}, 277 (1998).
\bibitem{Touchette} H. Touchette and  S. Lloyd,  Phys. Rev. Lett. \textbf{84}, 1156 (2000).
\bibitem{Zurek1} W. H. Zurek, arXiv:quant-ph/0301076 (2003).
\bibitem{Kieu}  T. D. Kieu,  Phys. Rev. Lett. \textbf{93}, 140403 (2004).
\bibitem{Allahverdyan} A.E. Allahverdyan, R. Balian, Th.M. Nieuwenhuizen, J. Mod. Optics, \textbf{51}, 2703 (2004).
\bibitem{Touchette2} H. Touchette and  S. Lloyd, Physica A \textbf{331}, 140 (2004).
\bibitem{Quan}  H. T. Quan, Y. D. Wang, Y-x. Liu, C. P. Sun, and F. Nori, Phys. Rev. Lett. \textbf{97}, 180402 (2006).
\bibitem{Cao1} F. J. Cao, L. Dinis, J. M. R. Parrondo, Phys. Rev. Lett. \textbf{93}, 040603 (2004).
\bibitem{Kim0} K. H. Kim and H. Qian, Phys. Rev. Lett. \textbf{93}, 120602 (2004).
\bibitem{Kim} K. H. Kim and H. Qian, Phys. Rev. E \textbf{75}, 022102 (2007).
\bibitem{Lopez} B. J. Lopez, N. J. Kuwada, E. M. Craig, B. R. Long, and H. Linke, Phys. Rev. Lett. \textbf{101}, 220601  (2008).
\bibitem{Allahverdyan2} A. E. Allahverdyan and D. B. Saakian,  Europhys Lett. \textbf{81}, 30003 (2008).
\bibitem{Sagawa-Ueda1} T. Sagawa and  M. Ueda,  Phys. Rev. Lett. \textbf{100}, 080403 (2008).
\bibitem{Jacobs} K. Jacobs, Phys. Rev. A \textbf{80}, 012322 (2009).
\bibitem{Cao2} F. J. Cao and M. Feito, Phys. Rev. E \textbf{79}, 041118 (2009).
\bibitem{Touchette3} F.J. Cao, M. Feito, and H. Touchette, Physica A \textbf{388}, 113 (2009).
\bibitem{Sagawa-Ueda3} T. Sagawa and M. Ueda, Phys. Rev. Lett. \textbf{104}, 090602 (2010).
\bibitem{Ponmurugan} M. Ponmurugan, Phys. Rev. E \textbf{82}, 031129 (2010).
\bibitem{Suzuki}  Y. Fujitani and H. Suzuki, J. Phys. Soc. Jpn. \textbf{79}, 104003 (2010). 
\bibitem{Horowitz1} J. M. Horowitz and S. Vaikuntanathan, Phys. Rev. E \textbf{82}, 061120 (2010).
\bibitem{Toyabe3} S. Toyabe, T. Sagawa, M. Ueda, E. Muneyuki, and M. Sano, Nature Physics \textbf{6}, 988 (2010).
\bibitem{SWKim} S. W. Kim, T.  Sagawa, S. De Liberato, and M. Ueda, Phys. Rev. Lett. \textbf{106}, 070401 (2011).
\bibitem{Morikuni} Y. Morikuni and H. Tasaki, J. Stat. Phys. \textbf{143}, 1 (2011).
\bibitem{Ito} S. Ito and M. Sano, Phys. Rev. E \textbf{84}, 021123 (2011).
\bibitem{Horowitz2} J. M. Horowitz and J. M. R. Parrondo, Europhys Lett. \textbf{95}, 10005 (2011).
\bibitem{Abreu} D. Abreu and U. Seifert, Europhys Lett. \textbf{94}, 10001 (2011).
\bibitem{Jarzynski5} S. Vaikuntanathan and C. Jarzynski, Phys. Rev. E \textbf{83}, 061120 (2011).
\bibitem{Sagawa} T. Sagawa, J. Phys.: Conf. Ser. \textbf{297}, 012015 (2011).
\bibitem{Dong} H. Dong, D. Z. Xu, C. Y. Cai, and C. P. Sun, Phys. Rev. E \textbf{83}, 061108 (2011). 
\bibitem{Pekola} D. V. Averin, M. M\"{o}tt\"{o}nen, and J. P. Pekola, Phys. Rev. B \textbf{84}, 245448 (2011).
\bibitem{Horowitz3} J. M. Horowitz and J. M. R. Parrondo, New J. Phys. \textbf{13}, 123019 (2011). 
\bibitem{Granger} L. Granger and H. Kantz, Phys. Rev. E \textbf{84}, 061110 (2011). 
\bibitem{Lahiri} S. Lahiri, S. Rana, and A. M. Jayannavar, \textit{J. Phys. A: Math. Theor.} \textbf{45}, 065002 (2012).
\bibitem{Lu} Y. Lu and G. L. Long, Phys. Rev. E \textbf{85}, 011125 (2012).
\bibitem{Sagawa-Ueda4} T. Sagawa and M. Ueda, Phys. Rev. E \textbf{85}, 021104 (2012).


\bibitem{Szilard}  L. Szilard, Z. Phys. \textbf{53}, 840 (1929).


\bibitem{Cover-Thomas} T. M. Cover and J. A. Thomas, ``\textit{Elements of Information Theory}'' (John Wiley and Sons, New York, 1991).
\bibitem{Shannon}  C. Shannon, Bell System Technical Journal \textbf{27}, 379-423 and 623-656 (1948).

\bibitem{Groenewold} H. J. Groenewold, Int. J. Theor. Phys. \textbf{4}, 327 (1971).
\bibitem{Ozawa} M. Ozawa, J. Math. Phys. \textbf{27}, 759 (1986).


\bibitem{Vale} R. D. Vale and F. Oosawa, Adv. Biophys. \textbf{26}, 97 (1990).
\bibitem{Prost} F. Julicher, A. Ajdari, and J. Prost,  Rev. Mod. Phys. \textbf{69}, 1269 (1997). 
\bibitem{Parrondo2} J. M. R. Parrondo and B. J. De Cisneros, Appl. Phys. A \textbf{75},179 (2002).
\bibitem{Reimann} P. Reimann. Phys. Rept. \textbf{361}, 57 (2002).
\bibitem{Hanggi2}  P. H\"{a}nggi and F. Marchesoni, Rev. Mod. Phys. \textbf{81}, 387 (2009).
\bibitem{Bier} M. Bier, Biosystems \textbf{88}, 201 (2007). 
\bibitem{Schliwa}  M. Schliwa and G. Woehlke, Nature \textbf{422}, 759 (2003).


\bibitem{Schlichting} H. J. Schlichting  and V. Nordmeier, Math. Naturwiss. Unterr. \textbf{49}, 323 (1996).
\bibitem{Weele} K. van der Weele, D. van der Meer, M. Versluis and D. Lohse, Europhys. Lett. \textbf{53}, 328 (2001).
%Hysteretic clustering in granular gas.
\bibitem{Serreli}  V. Serreli, C-F. Lee, E. R. Kay, and D. A. Leigh, Nature \textbf{445}, 523 (2007).
\bibitem{Raizen} G. N. Price, S. T. Bannerman, K. Viering, E. Narevicius, and M. G. Raizen, Phys. Rev. Lett. \textbf{100}, 093004 (2008).


\bibitem{Millonas} M. M. Millonas, Phys. Rev. Lett. \textbf{74}, 10 (1995).
\bibitem{Jayannavar} A. M. Jayannavar, Phys. Rev. E \textbf{53}, 2957 (1996).
%Simple model for Maxwell's-demon-type information engine
\bibitem{Eggers} J. Eggers, Phys. Rev. Lett. \textbf{83}, 5322 (1999).
%Sand as Maxwell's Demon
\bibitem{Brey} J. J. Brey, F. Moreno, R. Garcia-Rojo, and M. J. Ruiz-Montero, Phys. Rev. E \textbf{65}, 011305 (2001).  
%Hydrodynamic Maxwell demon in granular systems 
\bibitem{Broeck} C. Van den Broeck, P. Meurs, and R. Kawai, New J. Phys. \textbf{7}, 10 (2005).
%From Maxwell demon to Brownian motor
\bibitem{Ruschhaupt} A. Ruschhaupt, J. G. Muga, and M. G. Raize, J. Phys. B: At. Mol. Opt. Phys. \textbf{39}, 3833 (2006).



\bibitem{Jarzynski1} C. Jarzynski, Phys. Rev. Lett.  \textbf{78}, 2690 (1997).
%\bibitem{Kurchan} J. Kurchan, J. Phys. A. Math. Gen. \textbf{31}, 3719 (1998).
\bibitem{Crooks1} G. E. Crooks, J. Stat. Phys. \textbf{90}, 1481 (1998).
\bibitem{Crooks2}  G. E.  Crooks, Phys. Rev. E \textbf{60}, 2721 (1999).
\bibitem{Jarzynski2} C. Jarzynski, J. Stat. Phys. \textbf{98}, 77 (2000).
\bibitem{Seifert} U. Seifert, Phys. Rev. Lett. \textbf{95}, 040602 (2005).
\bibitem{Kawai}  R. Kawai, J. M. R. Parrondo, and C. Van den Broeck,  Phys. Rev. Lett.  \textbf{98}, 080602 (2007).
\bibitem{Bustamante} C. Bustamante, J. Liphardt,  and F. Ritort, Physics Today, \textbf{58}, 43 (2005).
\bibitem{Liphardt} J. Liphardt \textit{et al.},  Science  \textbf{296}, 1832 (2002).
\bibitem{Collin}  D. Collin \textit{et al.},  Nature \textbf{437}, 231 (2005).

\bibitem{Schreiber} T. Schreiber, Phys. Rev. Lett. \textbf{85}, 461 (2000).


\bibitem{Brillouin} L. Brillouin,  J. Appl. Phys. \textbf{22}, 334 (1951).
\bibitem{Landauer} R. Landauer,   IBM J. Res. Dev. \textbf{5}, 183 (1961).
\bibitem{Bennett} C. H. Bennett,  Int. J. Theor. Phys. \textbf{21}, 905 (1982).
\bibitem{Zurek2} W. H. Zurek, Nature \textbf{341}, 119 (1989).
\bibitem{Zurek3} W. H. Zurek, Phys. Rev. A \textbf{40}, 4731 (1989).
\bibitem{Shizume} K. Shizume, Phys. Rev. E \textbf{52}, 3495 (1995).
%\bibitem{Landauer2} R. Landauer, Science \textbf{272}, 1914 (1996).
\bibitem{Goto} H. Matsueda, E. Goto, and K-F. Loe, RIMS K\^{o}ky\^{u}roku \textbf{1013}, 187 (1997).
\bibitem{Piechocinska} B. Piechocinska,  Phys. Rev. A \textbf{61}, 062314 (2000).
\bibitem{Bennett2}  C. H. Bennett, Stud. Hist. Phil. Mod. Phys. \textbf{34},  501 (2003).
\bibitem{Allahverdyan3} A. E. Allahverdyan and T.M. Nieuwenhuizen,  Phys. Rev. E \textbf{64}, 0561171 (2001).
\bibitem{Horhammer} C. Horhammer and H. Buttner, J. Stat. Phys. \textbf{133}, 1161 (2008).
\bibitem{Barkeshli}  M. M. Barkeshli, arXiv:cond-mat/0504323 (2005).
\bibitem{Norton}  J. D. Norton, Stud. Hist. Phil. Mod. Phys. \textbf{36}, 375 (2005).
\bibitem{Maroney} O. J. E. Maroney, Phys. Rev. E \textbf{79}, 031105 (2009).
\bibitem{Turgut} S. Turgut, Phys. Rev. E \textbf{79}, 041102 (2009).
\bibitem{Sagawa-Ueda2} T. Sagawa and M. Ueda, Phys. Rev. Lett. \textbf{102}, 250602 (2009); \textbf{106}, 189901(E) (2011).

\bibitem{Kay} E. R. Kay, D. A. Leigh, and F. Zerbetto, Angew. Chem. \textbf{46}, 72 (2007).

 

\end{thebibliography}
\end{document}